\documentclass[11pt]{article}
\usepackage{authblk}
\usepackage{graphicx}
\usepackage{amssymb}
\usepackage{amsmath}
\usepackage{hyperref}
\usepackage{natbib}
\usepackage[margin=1in]{geometry}
\usepackage{subcaption}
\usepackage{tikz}
\usetikzlibrary{calc,fit,arrows.meta,arrows,mindmap,external}
\usepackage{color}
\usepackage{epstopdf, epsfig}
\usepackage{url}
\usepackage{newunicodechar}
\usepackage{natbib}
\usepackage{tabularx}
\usepackage{bm}
\usepackage{bbm}
\usepackage{setspace}
\usepackage[T1]{fontenc}
\usepackage[geometry]{ifsym}
\usepackage{authblk}

\newcommand\Str{\mbox{St}}           
\newcommand\Rey{\mbox{Re}}           
\newcommand\VK{von K\'arm\'an }

\colorlet{revision1}{black}

\begin{document}
\title{Direct Numerical Simulations of the Swirling \VK Flow Using a Semi-implicit Moving Immersed Boundary Method}
\author{M. Houssem Kasbaoui\thanks{houssem.kasbaoui@asu.edu} }
\affil{
  School for Engineering of Matter, Transport and Energy, Arizona State University, Tempe, AZ 85281, USA
}
\author{Tejas Kulkarni} 
\author{Fabrizio Bisetti}
\affil{
  Department of Aerospace Engineering and Engineering Mechanics, The University of Texas at Austin, Austin, TX 78712, USA
}

\date{\today}

\maketitle

\begin{abstract}
  We present a novel moving immersed boundary method (IBM) and employ it in direct numerical simulations (DNS) of the closed-vessel swirling \VK flow in laminar and turbulent regimes. The IBM extends direct-forcing approaches by leveraging a time integration scheme, that embeds the immersed boundary forcing step within a semi-implicit iterative Crank-Nicolson scheme. The overall method is robust, stable, and yields excellent results in canonical cases with static and moving boundaries. The moving IBM allows us to reproduce the geometry and parameters of the swirling \VK flow experiments in (F. Ravelet, A. Chiffaudel, and F. Daviaud, JFM 601, 339 (2008)) on a Cartesian grid. In these DNS, the flow is driven by two-counter rotating impellers fitted with curved inertial stirrers. We analyze the transition from laminar to turbulent flow by increasing the rotation rate of the counter-rotating impellers to attain the four Reynolds numbers 90, 360, 2000, and 4000. In the laminar regime at Reynolds number 90 and 360, we observe flow features similar to those reported in the experiments and in particular, the appearance of a symmetry-breaking instability at Reynolds number 360. We observe transitional turbulence at Reynolds number 2000. Fully developed turbulence is achieved at Reynolds number 4000. Non-dimensional torque computed from simulations matches correlations from experimental data. The low Reynolds number symmetries, lost with increasing Reynolds number, are recovered in the mean flow in the fully developed turbulent regime, where we observe two tori symmetrical about the mid-height plane. We note that turbulent fluctuations in the central region of the device remain anisotropic even at the highest Reynolds number 4000, suggesting that isotropization requires significantly higher Reynolds numbers.

\end{abstract}

\section{Introduction}

  Engineering flows operated in closed vessels, such as internal combustion engines and stirred tank reactors, are often subject to high levels of shear and velocity fluctuations. In these flows, the interaction between moving surfaces and the flow controls macroscopic quantities such as mixing rates and power consumption \citep{bertrandPowerConsumptionPumping1980}. 
  In the present paper, we develop a moving immersed boundary (IB) strategy that enables the study of highly turbulent flows interacting with moving components. We validate the method in canonical cases then apply it in direct numerical simulations (DNS) of the inertially-driven swirling \VK flow, a closed vessel flow of fundamental and practical interest.
  We show that laminar and turbulent regimes of the swirling \VK flow can be reproduced successfully by DNS with our IB method and analyze the homogeneity and anisotropy of the flow in the fully developed turbulence regime.

  Owing to its fundamental nature, the swirling \VK flow received significant attention.
  In his pioneering work, Theodor \VK \citep{karmanUberLaminareUnd1921} considered the flow over an infinite disk rotating at a rate $\Omega$. Von K\'arm\'an noted that the flow is self-similar and that the Navier-Stokes equations may be reduced to a pair of non-linear ordinary differential equations. \citet{batchelorNOTECLASSSOLUTIONS1951} further generalized the analysis to include a second coaxial disk at a distance $H$. The solution to these equations is chaotic and offers key insights into the non-linearity of the Navier Stokes equations.
  The earlier work of \VK and Batchelor was followed by sustained research efforts to analyze the flow characteristics in various regimes (see review of \citet{zandbergenKarmanSwirlingFlows1987}).
  More recently, the case of counter-rotating finite disks of radius $R$ received significant attention. While the flow is characterized by symmetry at low Reynolds number $\Rey_\Omega$, several authors reported the appearance of symmetry-breaking hydrodynamic instabilities with increasing $\Rey_\Omega$ \citep{lopezInstabilityModeInteractions2002,noreRatioModeInteraction2003,noreSurveyInstabilityThresholds2004,cortetSusceptibilityDivergencePhase2011}. Here, the Reynolds number is defined as $\Rey_\Omega=\Omega R^2/\nu$, where $\nu$ is the kinematic viscosity.  The resulting flow structures are stable and persist for a wide range of intermediate Reynolds numbers before the onset of additional symmetry-breaking instabilities \citep{raveletBifurcationsGlobalesHydrodynamiques2005,raveletSupercriticalTransitionTurbulence2008}. At large Reynolds numbers, a turbulent shear layer forms between two stacked toroidal cells of size comparable to the disk diameter.

  From an experimental perspective, the case of counter-rotating disks is of particular interest as it produces high Reynolds number turbulence inside a closed and compact device. \citet{maurerStatisticsTurbulenceTwo1994} produced a turbulent \VK flow with Taylor micro-scale Reynolds number $\Rey_\lambda\sim2100$ in a device of disk radius and separation $R=3.2\,\mbox{cm}$ and $H=4.8\,\mbox{cm}$, respectively.  \citet{odierAdvectionMagneticField1998} achieved a macroscopic Reynolds number $\Rey_\Omega=\Omega R^2/\nu=O(10^6)$ with $R=H=10\,\mbox{cm}$. Curved inertial-stirrers mounted on the disks increase velocity fluctuations and are used to tune flow structures \citep{raveletExperimentalKarmanDynamo2005,raveletBifurcationsGlobalesHydrodynamiques2005,burnishevTorquePressureFluctuations2014}. Access to such high Reynolds number regimes enables studies of fundamental turbulence properties such as intermittency, energy dissipation, and the turbulence cascade \citep{monchauxFluctuationDissipationRelationsStatistical2008,debueDissipationIntermittencySingularities2018,dubrulleKolmogorovCascades2019,kuzzayGlobalVsLocal2015}.

  Further work on the fine scale structures of the \VK flow requires spatial and temporal resolutions for which numerical studies are in principle better suited. Unlike the vigorous experimental effort deployed so far, investigations of the \VK flow relying on direct numerical simulations remain scarce.
  Few studies resolved the flow around the blades \citep{kreuzahlerNumericalStudyImpellerdriven2014,noreNumericalSimulationKarman2018}. While the existence of numerous experimental data sets enables insightful comparisons between experiments and simulations, it remains to be shown that current numerical methods are  able to reproduce experimental findings. Addressing this issue requires a computational strategy that manages computational cost while ensuring an accurate representation of the flow near the impellers.

  Immersed boundary (IB) methods are a natural choice for simulations that involve complex geometries, such as the inertially-driven swirling \VK flow. These methods remove the cumbersome task of generating body-conformal meshes and enable the use of straightforward Cartesian grids  for the discretization of the volume occupied by the fluid. Various approaches are summarized in \citep{mittalImmersedBoundaryMethods2005}. The so-called direct-forcing IB method \citep{peskinFlowPatternsHeart1972} relies on a forcing term added to the right hand-side of the momentum equation in order to impose  no-slip boundary conditions. This class of methods is amenable to efficient discretization and can handle moving immersed boundaries robustly \citep{romaAdaptiveVersionImmersed1999,peskinImmersedBoundaryMethod2002,laiImmersedBoundaryMethod2000}. Since the original work of Peskin, various improvements have been proposed \textcolor{revision1}{\citep{fadlunCombinedImmersedBoundaryFiniteDifference2000,kimImmersedBoundaryFiniteVolumeMethod2001,balarasModelingComplexBoundaries2004,kimImmersedBoundaryMethod2006,nicolaouRobustDirectforcingImmersed2015,kangDNSBuoyancydominatedTurbulent2009a}}. In particular, \citet{uhlmannImmersedBoundaryMethod2005} proposed that the forcing be applied on Lagrangian points distributed on the surface of the immersed solid. This method is characterized by its robustness and stability. Variations of Uhlmann's Lagrangian direct-forcing method have been proposed \citep{yangEmbeddedboundaryFormulationLargeeddy2006,vanellaMovingleastsquaresReconstructionEmbeddedboundary2009}.

  In the present work, we conduct DNS of the swirling \VK flows using a novel moving IB method  derived from \citet{uhlmannImmersedBoundaryMethod2005}'s method. First, we show that the properties of the Lagrangian markers (position and size) can be obtained from a triangular tessellation of the IB surface. Second, we couple the IB forcing to the update of the velocity and pressure fields by means of operator-splitting within a semi-implicit iterative Crank-Nicolson scheme for the advancement of momentum in incompressible flows \citep{pierceProgressVariableApproachLargeEddy2001,pierceProgressvariableApproachLargeeddy2004,choiEffectsComputationalTime1994}. The overall scheme allows a rapid workflow, whereby a mesh of the \VK flow enclosure and impellers generated by a CAD software is loaded in a direct numerical simulation flow solver without further adjustment. Data generated with this approach is compared to the experiments of \citet{raveletSupercriticalTransitionTurbulence2008} in the laminar and turbulent regimes.

  The paper is organized as follows. The governing equations and numerical discretization are introduced in Section \ref{sec:governing_equations}. In Section \ref{sec:validation_cases}, we validate the method in three benchmark cases including static and moving IBs. Simulations of the inertially-driven swirling \VK flow are presented in Section \ref{sec:von_karman}. Two laminar cases at $\Rey_\Omega=90$ and $\Rey_\Omega=365$ are considered in Section \ref{sec:VK_laminar}. Two additional cases at $\Rey_\Omega=2000$ and $\Rey_\Omega=4000$ are considered in Sections \ref{sec:VK_turbulent}. Final remarks are given in Section \ref{sec:conclusion}.

\section{Equations and methods}
\label{sec:governing_equations}
\subsection{Governing equations}

  \begin{figure*}
    \centering
    \includegraphics[width=0.5\linewidth]{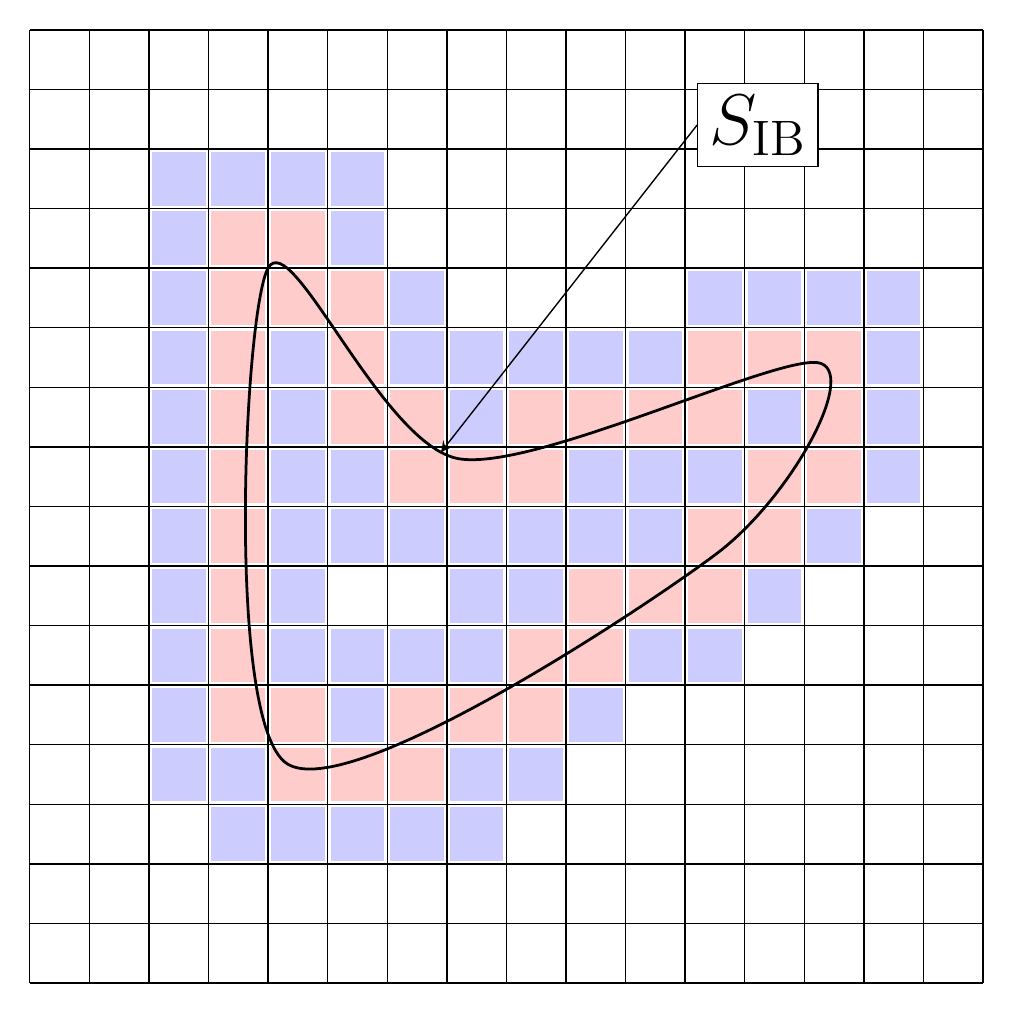}
    \caption{No-slip boundary conditions on an immersed solid are enforced using a forcing term applied to the right hand-side of the momentum equation. To maintain a sharp representation of the interface, the term has a compact support of three mesh widths applied to the cut-cells (red) and closest neighbors (blue). \label{fig_sketch_IB}}
  \end{figure*}

  Consider a solid with boundary surface $S_\mathrm{IB}$ immersed in an incompressible fluid of density $\rho$ and viscosity $\mu$.
  The fluid obeys mass and momentum conservation equations
  \begin{eqnarray}
    \nabla \cdot \bm{u}&=&0 \label{eq:continuity} \\
    \rho\left(\frac{\partial\bm{u}}{\partial t} + \bm{u}\cdot \nabla\bm{u}\right)&=& -\nabla p + \mu\nabla^2 \bm{u} +\bm{F}_\mathrm{IB},  \label{eq:NVS}
  \end{eqnarray}
  where $\bm{u}$ is the fluid velocity and $p$ is the pressure. In the direct-forcing approach \citep{peskinFlowPatternsHeart1972,peskinImmersedBoundaryMethod2002}, the IB forcing term
  \begin{equation}
    \bm{F}_\mathrm{IB}(\bm{x},t)= \iint_{\bm{y}\in S_\mathrm{IB}} \bm{f}_\mathrm{IB}(\bm{y},t)\delta(\bm{x}-\bm{y})dS\label{eq:FIB}
  \end{equation}
  enforces no-slip boundary conditions on the surface of the immersed solid. The field $\bm{f}_\mathrm{IB}$ represents the Lagrangian forcing at a location $\bm{y}$ belonging to the immersed surface $S_\mathrm{IB}$. Multiple immersed bodies are addressed by splitting $S_\mathrm{IB}$ into an arbitrary number of sets.

  In addition to (\ref{eq:continuity}) and (\ref{eq:NVS}), additional equations describing the motion of the solid may be added and coupled to the governing equations for the fluid. The IB forcing term (\ref{eq:FIB}) provides the coupling force between the immersed solid and the fluid. In the present work, we consider immersed solids with a prescribed rigid body motion.

  \subsection{Overview of the algorithm}
  \begin{figure*}
    \centering
    \includegraphics[height=0.5\textheight]{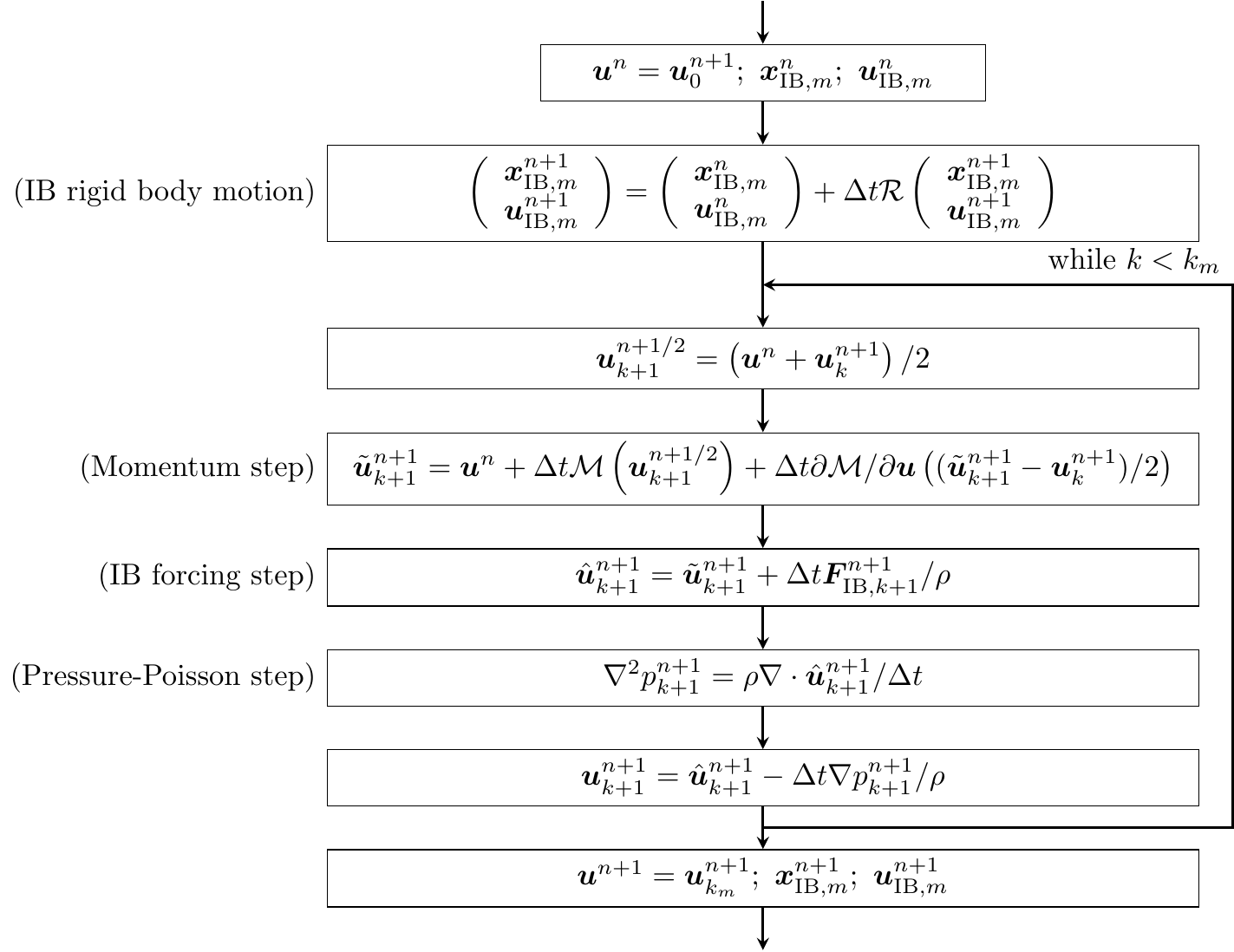}
    \caption{Algorithm flow chart showing the successive momentum, IB and pressure steps. The time integration relies on a semi-implicit iterative crank-Nicolson scheme and operator splitting.\label{fig:flow_chart}}
  \end{figure*}
  The governing equations are discretized and solved by the massively-parallel code NGA \citep{desjardinsHighOrderConservative2008}. The algorithm is shown in Fig \ref{fig:flow_chart}. The immersed boundary is discretized using a tessellation of $N$ triangular facets $S_m$, such that $S_\mathrm{IB}= \cup_{m=1}^N S_m$. At the beginning of each time step, the position of the immersed boundary is updated by moving the centroids of the triangles from their previous locations $\bm{x}^n_m$ to new positions $\bm{x}^{n+1}_m$ according to the prescribed rigid body motion.

  Next, the velocity field is updated while enforcing mass conservation. The time integration scheme for the momentum and pressure relies on the semi-implicit iterative Crank-Nicolson scheme introduced by \citep{akselvollLargeEddySimulation1995} and further developed in \citep{pierceProgressVariableApproachLargeEddy2001,pierceProgressvariableApproachLargeeddy2004}. We use an operator splitting approach to update the momentum and pressure, while considering the effects of the immersed solids, in three consecutive updates.

  Consider the $(k+1)^\mathrm{th}$ sub-iteration. First, we perform a conventional momentum update, where the IB forcing term and pressure term are omitted. The update reads
  \begin{equation}
   \tilde{\bm{u}}^{n+1}_{k+1}= \bm{u}^n +\Delta t \mathcal{M}(\bm{u}^{n+1/2}_{k+1})+\Delta t \frac{\partial \mathcal{M}}{\partial \bm{u}}\left(\frac{ \tilde{\bm{u}}^{n+1}_{k+1}-\bm{u}^{n+1}_{k} }{2}\right).\label{eq:mom_update}
  \end{equation}

  In the above, the mid-step velocity is $\bm{u}^{n+1/2}_{k+1}= (\bm{u}^n+\bm{u}^{n+1}_k)/2$ and $\mathcal{M}$ is the operator comprising both convective and viscous terms
  \begin{equation}
    \mathcal{M}(\bm{u})=-\nabla\cdot(\bm{u}\bm{u})+\frac{\mu}{\rho}\nabla^2\bm{u}.
  \end{equation}
  The Jacobian $\partial M/\partial \bm{u}$ in equation (\ref{eq:mom_update}) allows the treatment of the non-linearity with a Newton-Raphson method \citep{pierceProgressVariableApproachLargeEddy2001}. The momentum equation is solved with the approximate factorization technique of \citet{choiEffectsComputationalTime1994} based on the Alternating Direction Implicit (ADI) method. The method conserves mass, momentum and kinetic energy discretely \citep{pierceProgressVariableApproachLargeEddy2001,desjardinsHighOrderConservative2008,choiEffectsComputationalTime1994}.

   Next, the velocity is updated by applying the IB forcing
  \begin{equation}
   \hat{\bm{u}}^{n+1}_{k+1}= \tilde{\bm{u}}^{n+1}_{k+1} +\Delta t \bm{F}^{n+1}_{\mathrm{IB},k+1}/\rho. \label{eq:IB_step1}
  \end{equation}
  Lastly, a pressure-projection step is performed to enforce continuity by solving a Poisson equation and later correcting the velocity
  \begin{eqnarray}
   \nabla^2 p^{n+1}_{k+1}&=& \rho \frac{\nabla\cdot\hat{\bm{u}}^{n+1}_{k+1}}{\Delta t},\\
   \bm{u}_{k+1}^{n+1}&=&\hat{\bm{u}}^{n+1}_{k+1}-\frac{\Delta t}{\rho}\nabla p^{n+1}_{k+1}.
  \end{eqnarray}
  These sub-iterations are embedded within the iterative Crank-Nicolson loop. Typically, two to three subiterations per time step are used \citep{pierceProgressVariableApproachLargeEddy2001}. Note that if the Jacobian term is omitted and only two sub-iterations are retained, the time discretization becomes equivalent to an explicit second order Runge-Kutta scheme.

  \subsection{Treatment of the immersed boundaries}
  We now focus on the discretization of the forcing term in equation (\ref{eq:IB_step1}). Since $S_\mathrm{IB}= \cup_{m=1}^N S_m$, the forcing can be written as the sum of discrete contributions
  \begin{equation}
    \bm{F}^{n+1}_{\mathrm{IB},k+1}(\bm{x})= \sum_{m=1}^N \iint_{\bm{y} \in S_m} \bm{f}^{n+1}_{\mathrm{IB},k+1}(\bm{y})\delta_h(\bm{x}-\bm{y})dS,
  \end{equation}
  where, in the actual implementation, the Dirac delta is replaced by a regularized delta of finite width $h$ \citep{peskinImmersedBoundaryMethod2002}. \textcolor{revision1}{Note that, in this approach, internal cells are not forced}. The integrals on the facets are approximated to second-order accuracy using the mid-point rule
  \begin{equation}
    \bm{F}^{n+1}_{\mathrm{IB},k+1}(\bm{x})= \sum_{m=1}^N \bm{f}^{n+1}_{m,k+1}\delta_h(\bm{x}-\bm{x}_m)A_m,
  \end{equation}
  where $A_m$ is the surface area of facet $S_m$ and $\bm{f}^{n+1}_{m,k+1}$ is the Lagrangian IB forcing at the centroid $\bm{x}_m^{n+1}$. Following \citet{uhlmannImmersedBoundaryMethod2005},  no-slip boundary conditions are enforced on the immersed surface $S_\mathrm{IB}$ by ensuring that the fluid velocity equals the IB velocity $\bm{u}_{\mathrm{IB},m}$ at the centroid $\bm{x}_m$. This yields the following Lagrangian forcing terms
  \begin{equation}
    \bm{f}^{n+1}_{m,k+1}= \rho h\frac{ \bm{u}^{n+1}_{\mathrm{IB},m}-\tilde{\bm{u}}^{n+1}_{k+1}(\bm{x}_m)}{\Delta t}.
  \end{equation}
  Thus, the resulting Eulerian forcing term in (\ref{eq:IB_step1}) reads
  \begin{equation}
    \bm{F}^{n+1}_{\mathrm{IB},k+1} =\sum_{m=1}^N \rho\left(\frac{ \bm{u}^{n+1}_{\mathrm{IB},m}-\tilde{\bm{u}}^{n+1}_{k+1}(\bm{x}_m)}{\Delta t}\right)\delta_h(\bm{x}-\bm{x}_m)hA_m. \label{eq:discrete_FIB}
  \end{equation}

  The fluid velocity at the centroids is obtained by interpolation from  neighboring nodes on the Eulerian grid with $\delta_h$ as interpolation kernel. Here, we use the regularized Dirac delta proposed by \citet{romaAdaptiveVersionImmersed1999}, which has a compact support of width $h$. By choosing $h=3\Delta x$, where $\Delta x$ is the homogeneous mesh spacing, we ensure a sharp representation of the IB and efficient summation in (\ref{eq:discrete_FIB}).


\section{Validation cases}
\label{sec:validation_cases}
 In this section, we evaluate the accuracy and performance of the IB method against experimental and numerical data in canonical laminar flows. Three cases are discussed: the flow around a static cylinder placed  \textcolor{revision1}{asymmetrically} in a channel, the flow around a static cylinder in free stream, and the flow around a transversely oscillating cylinder.

\subsection{Static cylinder placed asymmetrically in a channel}
\begin{figure*}
  \centering
  \begin{subfigure}{\linewidth}
    \centering
     \includegraphics[width=0.80\linewidth]{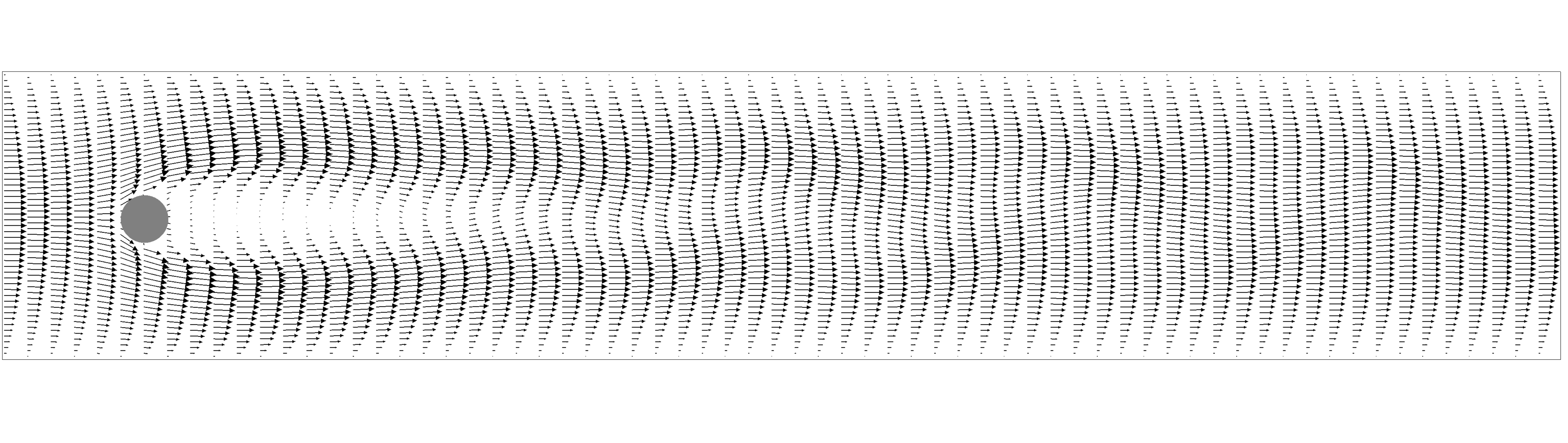}
     \caption{Vortex shedding in the wake of an asymmetrically placed cylinder. \label{fig:asym_cylinder}}
  \end{subfigure}\\
  \begin{subfigure}{0.45\linewidth}
    \centering
    \includegraphics[width=\linewidth]{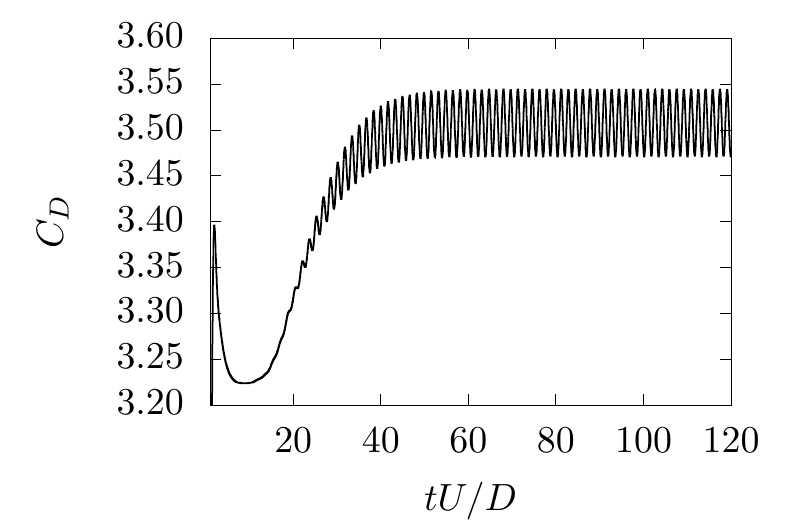}
    \caption{Drag coefficient \label{fig:asym_cylinder_drag}}
  \end{subfigure}
  \begin{subfigure}{0.45\linewidth}
    \centering
    \includegraphics[width=\linewidth]{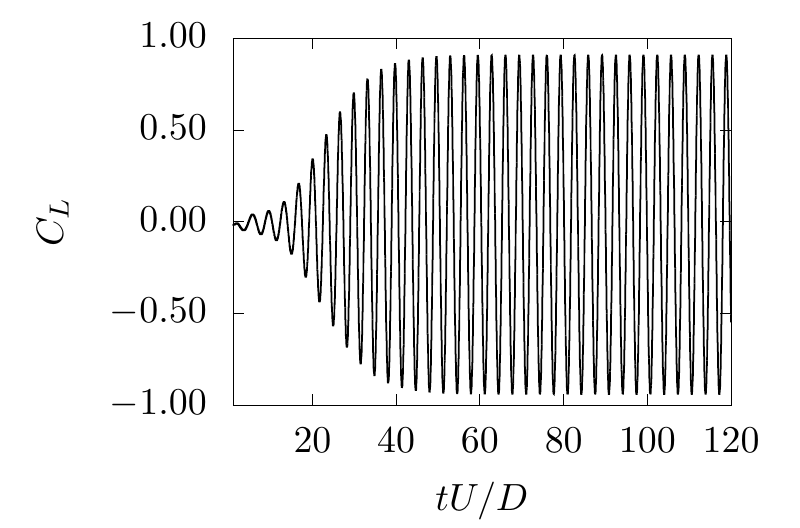}
    \caption{Lift coefficient \label{fig:asym_cylinder_lift}}
  \end{subfigure}
  \caption{Flow past an asymmetrically placed cylinder at $\Rey=100$. Data for $D/\Delta x= 48.8$ and $\mathrm{CFL}_\mathrm{max}\sim0.25$.}
\end{figure*}

\begin{table*}
  \caption{Strouhal number, drag and lift coefficients for the case of a cylinder placed asymmetrically in a channel at $\Rey=100$. \label{tab:asym_cylinder}}
  \newcolumntype{b}{X}
  \newcolumntype{s}{>{\hsize=.5\hsize}X}
  \def\refa{\citet{schaferBenchmarkComputationsLaminar1996}}
  \begin{tabularx}{\linewidth}{bssss}\hline
             & $D/\Delta x$& \Str         & $C_{D,max}$   & $C_{L,max}$ \\\hline\hline
    present  & 24.4        &  0.308       &  3.544        &  0.783      \\
    present  & 48.8        &  0.306       &  3.575        &  0.886      \\
    present  & 97.6        &  0.303       &  3.442        &  0.907      \\
    \refa    & -           & $0.3\pm0.005$& $3.23\pm0.01$ & $1.0\pm0.01$\\\hline
  \end{tabularx}
\end{table*}
We first consider the two-dimensional configuration in the benchmark flow of \citet{schaferBenchmarkComputationsLaminar1996}. A cylinder of diameter $D$ is placed in a channel of height $H=4.1D$ and length $L=22D$. The static cylinder is placed asymmetrically at $x=y=0.3$. A parabolic inflow with average velocity $U$ is prescribed at the inlet $x=0$. The fluid kinematic viscosity $\nu$ is such that $\Rey_D=UD/\nu=100$. Three spatial resolutions are considered where $D/\Delta x$ equals 24.4, 48.8 and 97.6, respectively. In all configurations, the maximum Courant–Friedrichs–Lewy number $\mathrm{CFL}$ is $\sim 0.25$.

The flow around the cylinder results in an oscillating wake, as shown in Fig. \ref{fig:asym_cylinder}.
Vortex shedding leads to fluctuating drag and lift coefficients as in Fig. \ref{fig:asym_cylinder_drag} and \ref{fig:asym_cylinder_lift}. Once a stationary state sets in after  $tU/D\sim200$, we collect statistics from the time histories of drag and lift forces.

 Comparison with the data in \citet{schaferBenchmarkComputationsLaminar1996} is shown in Tab. \ref{tab:asym_cylinder}. We report the Strouhal number, maximum drag coefficient and maximum lift coefficient for increasing resolution from $D/\Delta x = 24.4$ to $97.6$. For the case with the highest resolution, the shedding frequency $f_0$ yields a characteristic Strouhal number $\Str=f_0 D/U\sim0.303$ well within the range $0.295-0.305$ in \citep{schaferBenchmarkComputationsLaminar1996}. The maximum drag coefficient and maximum lift coefficient fall within 7\% and 0.6\% of the values reported in the literature, respectively.

\subsection{Static cylinder in uniform crossflow}
\begin{figure*}
  \centering
  \begin{subfigure}{\linewidth}
    \centering
    \includegraphics[width=0.80\linewidth]{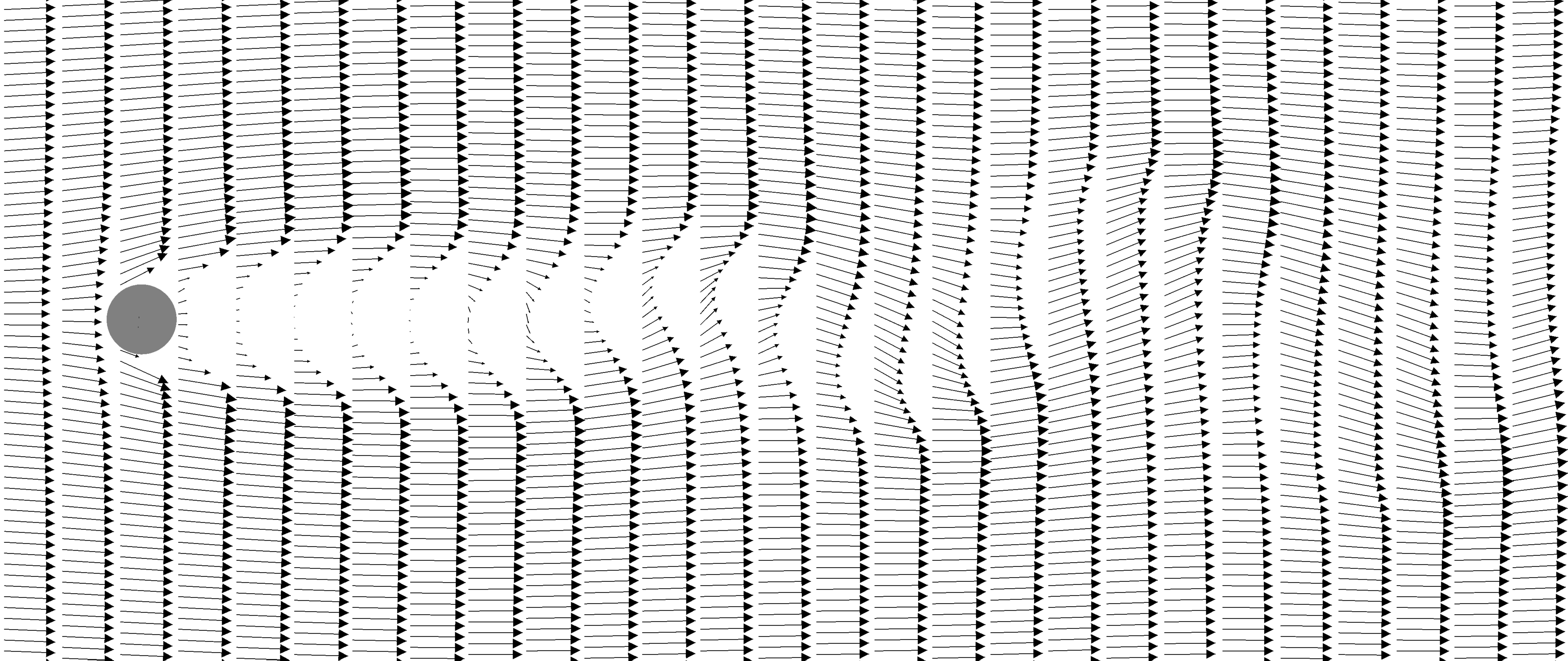}
     \caption{Vortex street behind an immersed cylinder at $\Rey_D=100$.\label{fig:static_cylinder}}
  \end{subfigure}\\
     \begin{subfigure}{0.45\linewidth}
       \centering
       \includegraphics[width=\linewidth]{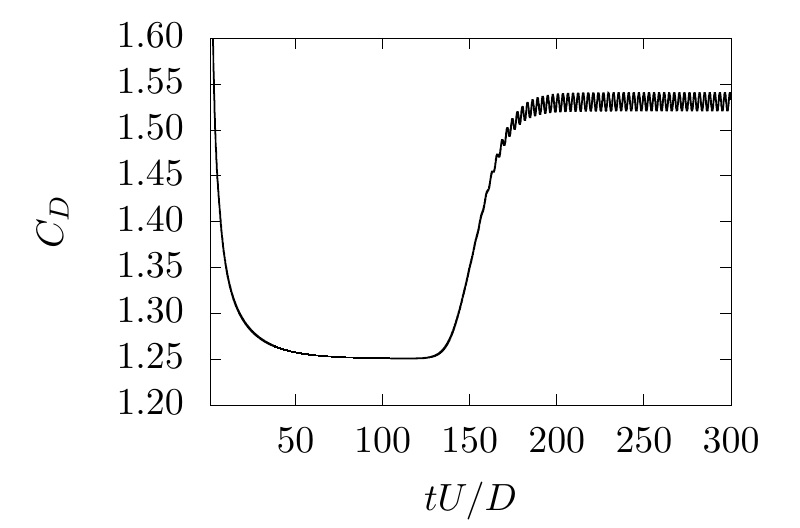}~
       \caption{Drag coefficient\label{fig:static_cylinder_drag}}
     \end{subfigure}
     \begin{subfigure}{0.45\linewidth}
       \centering
       \includegraphics[width=\linewidth]{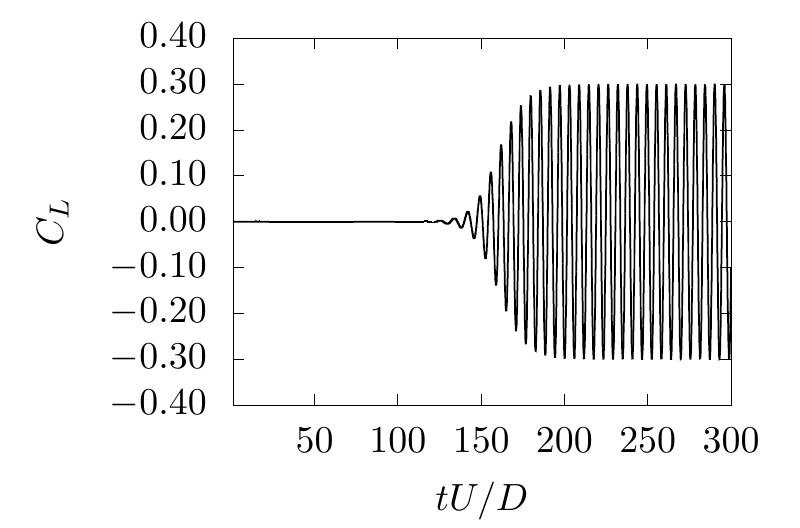}
       \caption{Lift coefficient\label{fig:static_cylinder_lift}}
     \end{subfigure}
     \caption{Drag and lift coefficients over an immersed cylinder in a uniform free stream at $\Rey_D=100$. Data for $D/\Delta x= 48.8$ and $\mathrm{CFL}\sim0.25$.}
\end{figure*}

  \begin{table}
    \caption{Strouhal number, drag and lift coefficients for the case of a cylinder in free stream at $\Rey=100$.  \label{tab:static_cylinder}}
    \newcolumntype{b}{X}
    \newcolumntype{s}{>{\hsize=.5\hsize}X}
    \def\refa{\citet{liuPreconditionedMultigridMethods1998}}
    \def\refb{\citet{williamsonObliqueParallelModes1989}}
    \begin{tabularx}{\linewidth}{bsssss}\hline
                         & $D/\Delta x$  &  \Str  & $\bar{C}_D$ & $C'_D$   & $C'_L$     \\\hline\hline
      Present            & 24.4          &  0.167 & 1.500       & 0.004    & 0.250      \\
      Present            & 48.8          &  0.167 & 1.526       & 0.005    & 0.289      \\
      Present            & 97.6          &  0.167 & 1.531       & 0.007    & 0.299      \\
      \refa              & --            &  0.165 & 1.350       & 0.012    & 0.339      \\
      \refb              & --            &  0.164 & --          & --       & --         \\\hline
    \end{tabularx}
  \end{table}

  Next, we consider a static cylinder of diameter $D=0.3$ placed in free stream with uniform inlet velocity. The computational domain has a size $26D\times26D$. The cylinder is located at $x_c=6D$ and $y_c=4D$ from the bottom left corner.  A uniform free-stream velocity $u_\infty=1$ is prescribed at the left inlet boundary, and convective outflow conditions are applied to the remaining boundaries. The Reynolds number is $\Rey_D=u_\infty D/\nu=100$. The domain is discretized on a uniform grid of size $128^2$, $256^2$ or $512^2$. The resulting resolution is $D/\Delta x=24.4$, 48.8 and 97.6. Note that the timestep $\Delta t$ is also adjusted to maintain $\mathrm{CFL}$ approximately constant at $0.25$.

  Figure \ref{fig:static_cylinder} shows the vortex street created by the immersed cylinder. The vortices are shed from the top and bottom sides of the cylinder at a natural frequency $f_0$. We obtain a Strouhal number $\Str=f_0D/u_\infty=0.167$ sensitively close to $0.164$ and $0.165$ determined from the experiments of \citet{williamsonObliqueParallelModes1989} and body-fitted simulations of \citet{liuPreconditionedMultigridMethods1998},respectively.

  Figure \ref{fig:static_cylinder_drag} and \ref{fig:static_cylinder_lift} show the time history of drag and lift coefficients. For the runs with highest resolution $D/\Delta x=97.6$, we find an average $C_D=1.531$ and a root mean square (rms) fluctuation $C'_D=0.007$. The mean lift coefficient is vanishingly small, as expected, while the fluctuation is $C'_L=0.299$. These values are compared with those in \citep{liuPreconditionedMultigridMethods1998} and shown in Tab \ref{tab:static_cylinder}.  We note that there is an over-prediction of the drag coefficient by 13\% and under-prediction of the rms lift coefficient by 12\%. This behavior is similar to the observations of \citet{uhlmannImmersedBoundaryMethod2005}, from which the method is derived.

\subsection{Cylinder oscillating transversely in free stream}

\begin{figure*}
  \centering
  \begin{subfigure}{0.45\linewidth}
    \includegraphics[width=\linewidth]{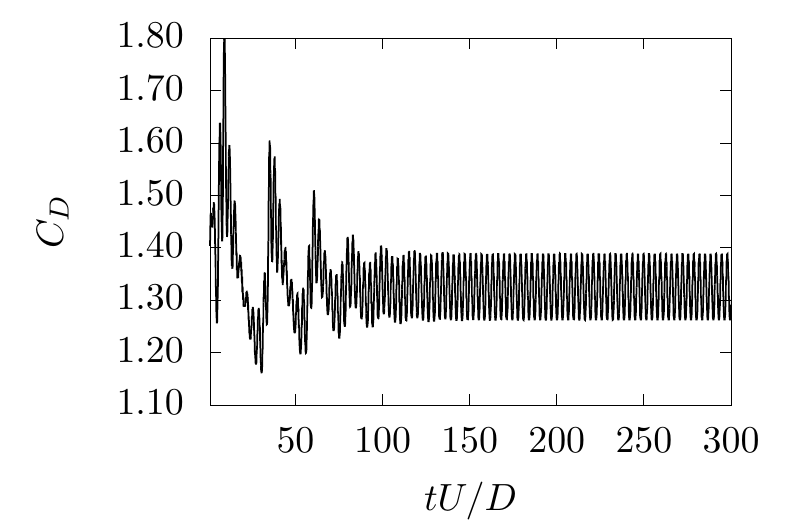}
    \caption{\label{fig:osci2_1}}
  \end{subfigure}
  \begin{subfigure}{0.45\linewidth}
    \includegraphics[width=\linewidth]{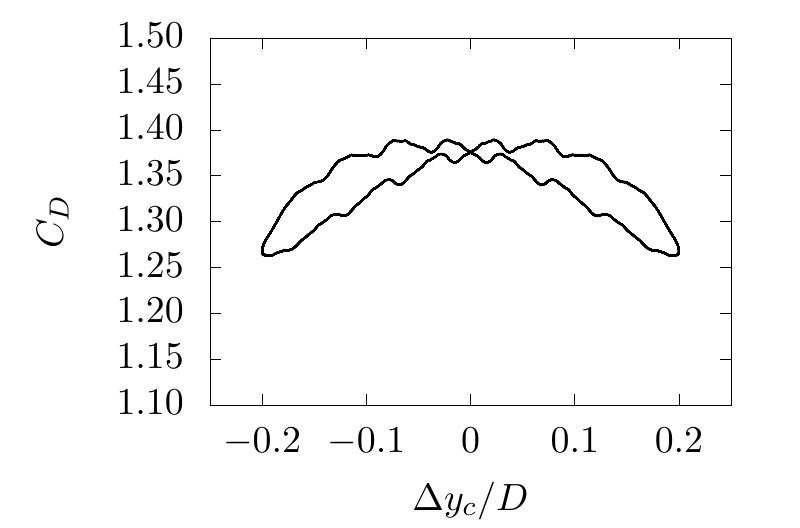}
    \caption{\label{fig:osci2_2}}
  \end{subfigure}
  \caption{Drag coefficient for a transversely oscillating cylinder in free stream at $\Rey=185$. Data for $D/\Delta x= 48.8$ and $\Delta t D/U_\mathrm{max}=0.0025$\label{fig:osci2}}
\end{figure*}

  \begin{table}
    \caption{Strouhal number, drag and lift coefficients for the case of a cylinder oscillating transversely at $\Rey=185$ and $f_e/f_0=0.8$. \label{tab:osci2_cylinder}}
    \newcolumntype{b}{X}
    \newcolumntype{s}{>{\hsize=.5\hsize}X}
    \def\refa{\citet{luCalculationTimingVortex1996}}
    \def\timestep{$\Delta t U_\mathrm{max}/D$}
    \begin{tabularx}{\linewidth}{bsssss}\hline
                         & \timestep  & $\bar{C}_D$ & $C'_D$    \\\hline\hline
      Present            & 0.0100     & 1.490       & 0.048     \\
      Present            & 0.0050     & 1.394       & 0.047     \\
      Present            & 0.0025     & 1.323       & 0.044     \\
      \refa              & --         & 1.25        & --        \\\hline
    \end{tabularx}
  \end{table}
The configuration described in the previous section is now modified to allow oscillations of the cylinder. The latter moves transversely whereby the displacement of the center is given by $\Delta y_c=0.2D\sin(2\pi f_et)$. The forcing oscillation frequency is $f_e=0.8f_0$, where $f_0$ is the natural shedding frequency for a fixed cylinder at Reynolds number $\Rey=185$. These parameters follow the simulations in \citep{luCalculationTimingVortex1996} using a body-fitted method. For this case, we maintain a fixed spatial resolution at $512\times 512$, giving a ratio $D/\Delta x=48.8$, while the timestep $\Delta t$ is set at $0.01U_\mathrm{max}/D$, $0.005U_\mathrm{max}/D$, or $0.0025U_\mathrm{max}/D$. The corresponding $\mathrm{CFL}$ is 0.5,0.25, and 0.125, respectively.

Figure \ref{fig:osci2} shows the evolution of the drag coefficient for the case with $\Delta t=0.0025 U_\mathrm{max}/D$. As seen in \ref{fig:osci2_1}, the drag coefficient reaches a stationary state after approximately $130 D/U$. For the case where $\Delta t=0.0025 U/D$, the average drag coefficient (see Tab. \ref{tab:osci2_cylinder}) is within 6\% of the value reported by \citet{luCalculationTimingVortex1996}. The drag curve plotted as a function of displacement in \ref{fig:osci2_2} follows a figure eight shape similar to the one found in \citep{uhlmannImmersedBoundaryMethod2005}. We note the presence of spurious oscillations in  Fig. \ref{fig:osci2_2} that increase the rms drag coefficient fluctuations. As argued in \citep{uhlmannImmersedBoundaryMethod2005}, these spurious oscillations can be reduced with larger discrete Dirac delta support than considered here.

\section{The swirling \VK flow}
\label{sec:von_karman}

  \begin{figure*}
    \centering
     \includegraphics[width=0.9\linewidth]{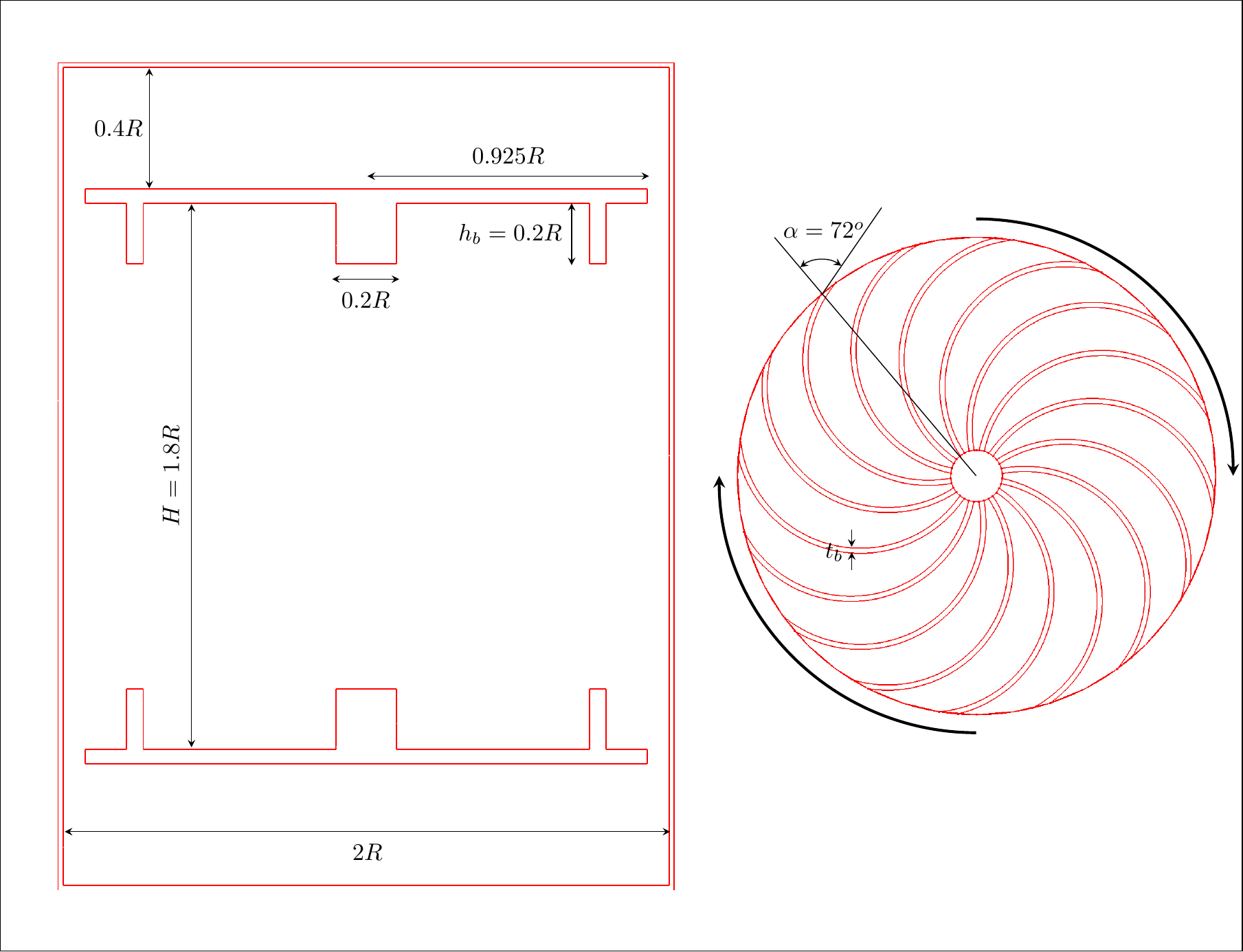}
     \caption{Geometry of the device. The top and bottom disks rotate in opposite directions at a constant rotation rate. The impeller design corresponds to the configuration TM60 analyzed in \citep{raveletSupercriticalTransitionTurbulence2008}.\label{fig:VK_geometry}}
  \end{figure*}

  We now apply the immersed boundary method described in Section \ref{sec:governing_equations} to simulations of the swirling \VK flow.

  The \VK flow considered in our work is generated in a closed cylindrical vessel between two counter-rotating disks fitted with curved blades as shown in Fig. \ref{fig:VK_geometry}. The numerical setup is a reproduction of the experimental apparatus analyzed by \citet{raveletSupercriticalTransitionTurbulence2008} given available information in \citep{raveletSupercriticalTransitionTurbulence2008,raveletBifurcationsGlobalesHydrodynamiques2005}.
The disks have radius equal to $0.925R$, where $R$ is the inner cylinder's radius, and separation $H=1.8R$. The impellers act as centrifugal pumps that ingest fluid along the centerline and expel it radially towards the cylindrical walls. Inertial stirring is aided by 16 blades mounted on the disks. The stirrers correspond to the TM60 design in \citep{raveletBifurcationsGlobalesHydrodynamiques2005}. They have a height $h_b= 0.2 R$, a thickness $t_b=0.02R $, and radius of curvature $C=R/(2\mbox{sin}\alpha)$, where the curvature angle is $\alpha=72^\circ$. All 16 blades are connected to a cylindrical hub of radius $0.1 R$ and height equal to that of the blades. Flow ejected towards the walls by the impellers may enter a recirculation regions behind the disks of height $0.4R$. 

  \begin{table}
  \caption{Simulation parameters for the five cases at $\Rey_\Omega=90$, 360, 2000 and 4000. \label{tab:VK_parameters}}

    \newcolumntype{b}{X}
    \newcolumntype{s}{>{\hsize=.3\hsize}X}
    \newcolumntype{r}{>{\hsize=.05\hsize}X}
    \def\a{\times10^{-4}}
    \begin{tabularx}{\linewidth}{bsssrss}\hline
                                    &                         &\multicolumn{2}{c}{laminar} & &\multicolumn{2}{c}{turbulent} \\\cline{3-4} \cline{6-7}
      Parameter                     & Symbol                  & case 1  & case 2  & & case 3  & case 4  \\\hline\hline
      Reynolds number $\Rey_\Omega$ & $\Omega R^2/\nu$        &  90     & 360     & & 2000    & 4000    \\
      Spatial resolution            & $R/\Delta x$            & 128.0   & 128.0   & & 128.0   & 256.0   \\
      Temporal resolution           & $\Delta t \Omega/2\pi$  & $3.2\a$ & $3.2\a$ & & $3.2\a$ & $1.6\a$ \\
      Kolmogorov scale              & $\eta/\Delta x$         & --      & --      & & 3.4     & 3.9     \\
      Taylor-micro scale Reynolds   & $\Rey_\lambda$          & --      & --      & & 8       & 40      \\\hline
    \end{tabularx}
  \end{table}

  We consider four simulations at Reynolds numbers $\Rey_\Omega=\Omega R^2/\nu=90$, 360, 2000 and 4000. A summary of the parameters is given in Tab. \ref{tab:VK_parameters}. The Reynolds number is adjusted by increasing the rotation rate of the disks.
  In all configurations, the grid is uniform with a constant mesh size $\Delta x$. The discretization of the IB surfaces is obtained from a Delaunay triangulation with an approximate element size $\sim \Delta x/2$. The surface of the \VK flow device consists of 3 sets: a cylindrical enclosure, top, and bottom impellers. The cylindrical enclosure is static. The top and bottom impellers rotate in opposite directions at a constant rate with the concave face of the blades pointing forward in the direction of motion. This is generally referred to as the (-) direction of rotation. Note that the opposite direction of rotation is not equivalent due to the asymmetry of the curved blades. No-slip boundary conditions imposed on the immersed boundaries constrain the flow to the interior of the swirling \VK flow device.

  The four configurations correspond to different regimes of the swirling \VK flow. As documented by \citet{raveletSupercriticalTransitionTurbulence2008}, the flow undergoes regime transitions from laminar to fully turbulent with increasing $\Rey_\Omega$. The transitions are characterized by gradual loss of symmetries. The flow at $\Rey_\Omega=90$ falls in the laminar regime described in \citep{raveletSupercriticalTransitionTurbulence2008}, where the flow is steady, axisymmetric, and symmetric about the mid-height plane. At $\Rey_\Omega=360$, the flow is steady and laminar and the symmetry about the mid-height plane is disrupted by an azimuthal wave of mode 2. \citet{raveletSupercriticalTransitionTurbulence2008} report transitional turbulence at $\Rey_\Omega=2000$ and fully developed turbulence past $\Rey_\Omega\sim3300$.  The mean flow is made of a shear layer centered on the mid-height plane formed between two toroidal structures.

 We maintain sufficient resolution for all four runs. \textcolor{revision1} {For the run at $\Rey_\Omega=4000$, 90 grid points lie between each blade at the tip of the rotating disks ($r=0.9R$). This ensures that the resolution is sufficient to capture the fluid stresses on the impellers, as shown in the grid convergence study in \ref{sec:appendix}. Good agreement with experimental torque data discussed below further supports that the fluid stresses on the impellers are captured adequately. The central region of the flow is also well resolved. The ratio of the Kolmogorov length scale to mesh width spacing is $\eta/\Delta x\simeq3.4$ and 3.9 for the runs at $\Rey_\Omega=2000$ and 4000, respectively. The Kolmogorov scale is computed at the center of the device from dissipation rate. }

  \subsection{Laminar regime}
  \label{sec:VK_laminar}
  
  \begin{figure*}
	\centering
    \includegraphics[width=\linewidth]{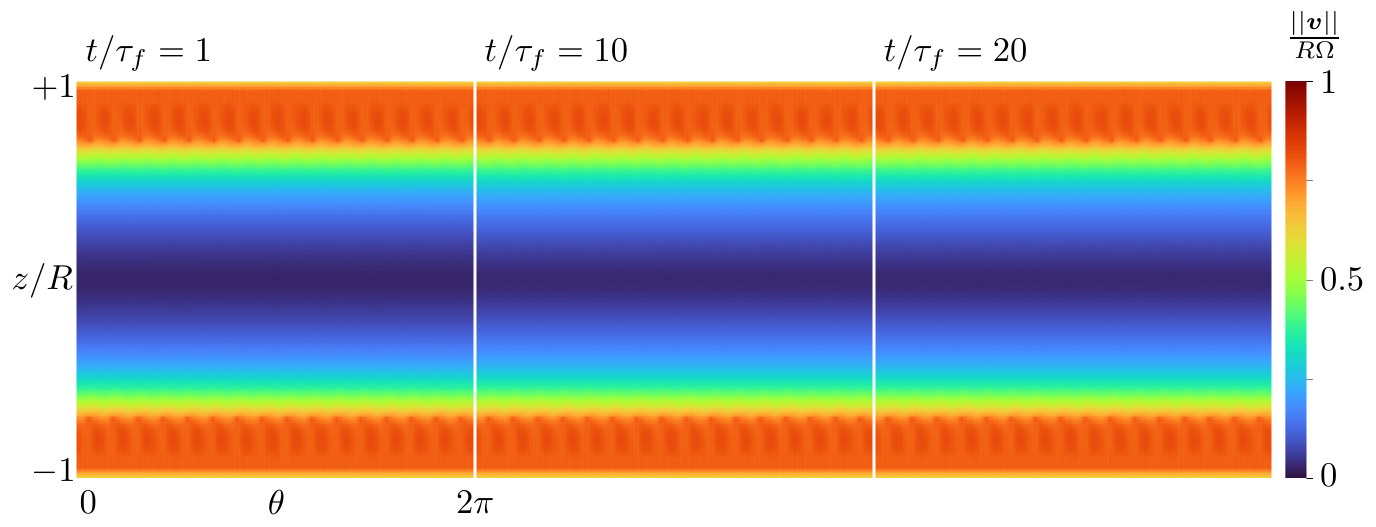}
    \caption{isocontours of the normalized velocity magnitude at  $\Rey_\Omega=90$ from $t/\tau_f=1$ to 20, where $\tau_f=\Omega/2\pi$ is the time it takes to complete a full revolution of the disks. The circumferential cut is taken at the radial distance $r/R=0.8$. At this Reynolds number, the flow is laminar, axisymmetric and planar symmetric about the mid-height plane. \label{fig:snap_90}}
  \end{figure*}
  
  \begin{figure*}
	\centering
      \includegraphics[width=0.7\linewidth]{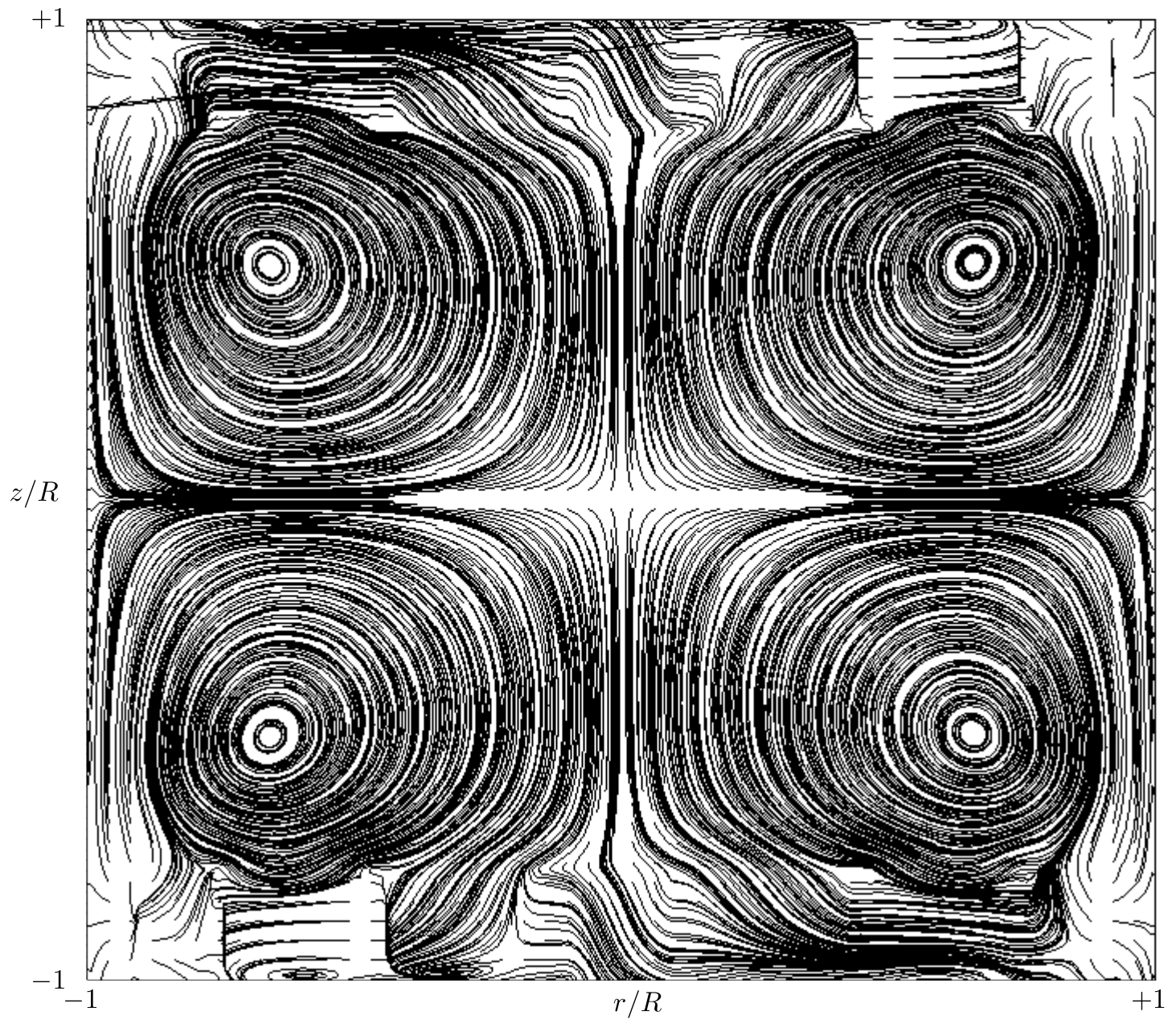}
    \caption{Streamlines of the velocity field in a vertical plane through the axis at $\Rey_\Omega=90$. The flow presents a shear layer formed in between two-toroidal cells similar to what has been reported experimentally in \citep{raveletSupercriticalTransitionTurbulence2008}. \label{fig:streamlines_90}}
  \end{figure*}
    
  We start with the flow at $\Rey_\Omega=90$. Fig. \ref{fig:snap_90} shows instantaneous isocontours of the velocity magnitude from $t\Omega/2\pi=1$ to 20, i.e., over 20 revolutions. The isocontours are visualized in a circumferential cut at the radial distance $r=0.8R$. It is apparent that a steady state is reached in less than one revolution of the impellers. Similarly to the experimental observation in \citep{raveletSupercriticalTransitionTurbulence2008}, the flow obtained in these simulations is axisymmetric and planar symmetric about the mid-height plane.

  Figure \ref{fig:streamlines_90} shows streamlines of the velocity field in a plane going through the axis. The figure shows the existence of a flat shear layer at the mid-height plane between two toroidal structures. These vortical structures are the result of the impellers drawing fluid towards their center and expelling it towards the cylinder walls. The fluid recirculates along the cylindrical enclosure's walls and returns to the center of the device at the mid-height plane thus creating the shear layer. The patterns observed from DNS are in excellent agreement with the structures seen in the photographs of \citet{raveletSupercriticalTransitionTurbulence2008} at the same Reynolds number.

  \begin{figure*}
  	\centering
    \includegraphics[width=\linewidth]{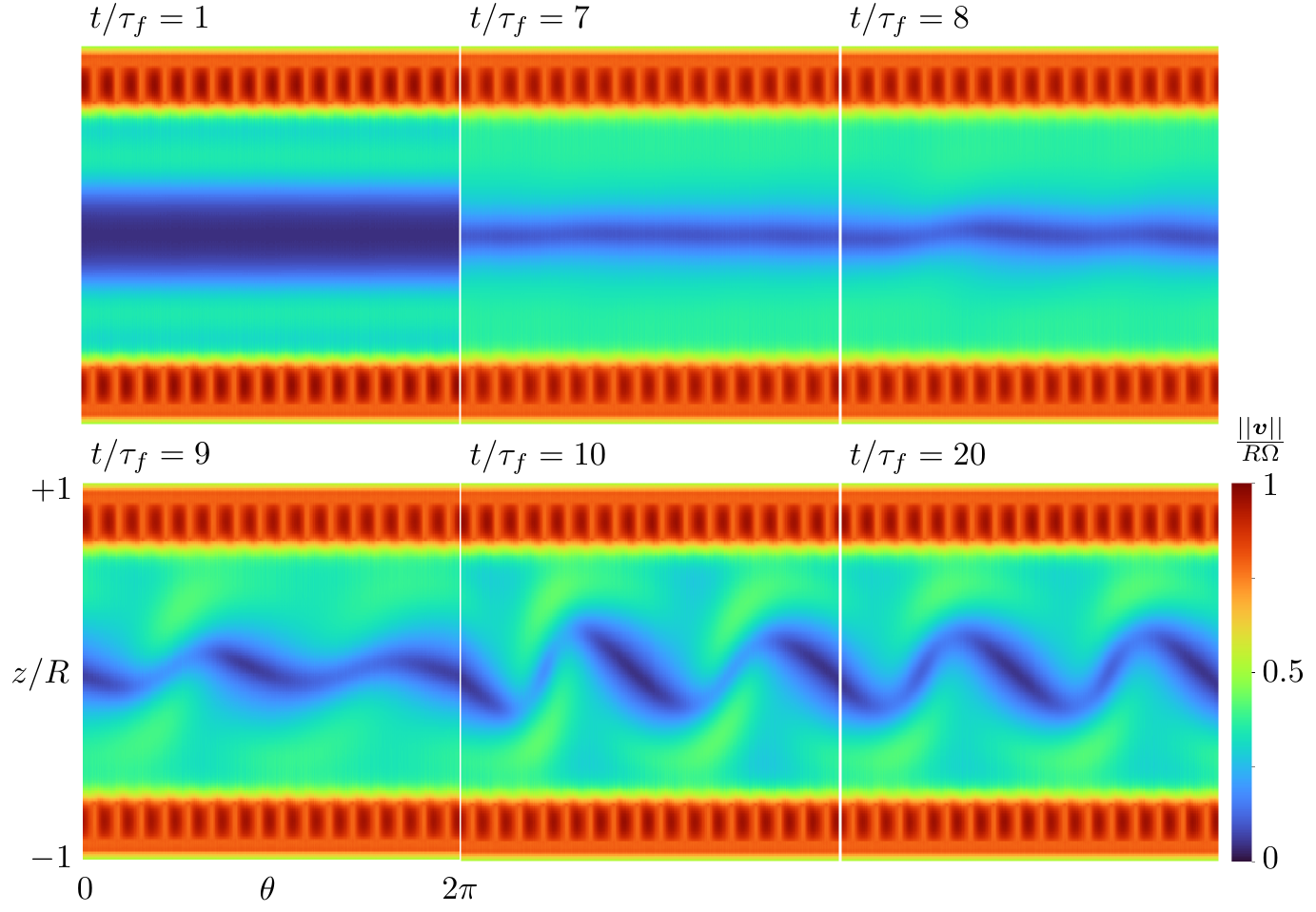}%
    \caption{Isocontours of the normalized velocity magnitude at $\Rey_\Omega=360$. The circumferential cut is taken at the radial distance $r/R=0.8$. A sudden transition occurs at $t\Omega /2\pi\sim 8$ leadings to the growth of an azimuthal velocity wave with mode $m=2$. \label{fig:snap_360}}
  \end{figure*}

  Isocontours of the normalized velocity magnitude of the flow at $\Rey_\Omega=360$ are shown in Fig. \ref{fig:snap_360}. 
  The simulations show that the flow starts with a symmetrical shear layer for $t\Omega/(2\pi)<8$. During this time, the shear layer becomes progressively thinner, which is indicative of increasing shear rate at the mid-height plane. A sudden instability of the shear layer breaks the axisymmetry at $t\Omega/(2\pi)\sim 8$ and leads to the emergence of an azimuthal velocity wave with mode $m=2$.   Unlike the lower Reynolds number case, the flow does not reach a steady state until $t\Omega/(2\pi)\sim 14$ when the mode $m=2$ stabilizes. 
  
  The physics revealed in our simulations are in accordance with the experimental observations of \citet{raveletSupercriticalTransitionTurbulence2008}. Long-exposure photographs of tracers in \citep{raveletSupercriticalTransitionTurbulence2008} at $\Rey_\Omega=345$ show the existence of an $m=2$ azimuthal mode. \citet{noreRatioModeInteraction2003} argue that the $m=2$ mode is due to a Kelvin-Helmholtz instability of the equatorial shear layer. \citet{raveletSupercriticalTransitionTurbulence2008} also note that the azimuthal mode in their experiments rotates slowly around the axis. They find that the shear layer completes a full revolution every 300 revolutions of the impellers. However, it is not clear what would cause the rotation of the mode in a preferred direction given that the top and bottom impellers are symmetrical. While we do not observe any noticeable rotation of the shear layer in the 20 rotations simulated here, ruling out the slow dynamics would require significantly longer integration time than we have considered.

  \subsection{Turbulent regime}
  \label{sec:VK_turbulent}
  \subsubsection{Regime identification and torque measurements: Validation against experimental data}

  \begin{figure*}
    \centering
    \includegraphics[width=\linewidth]{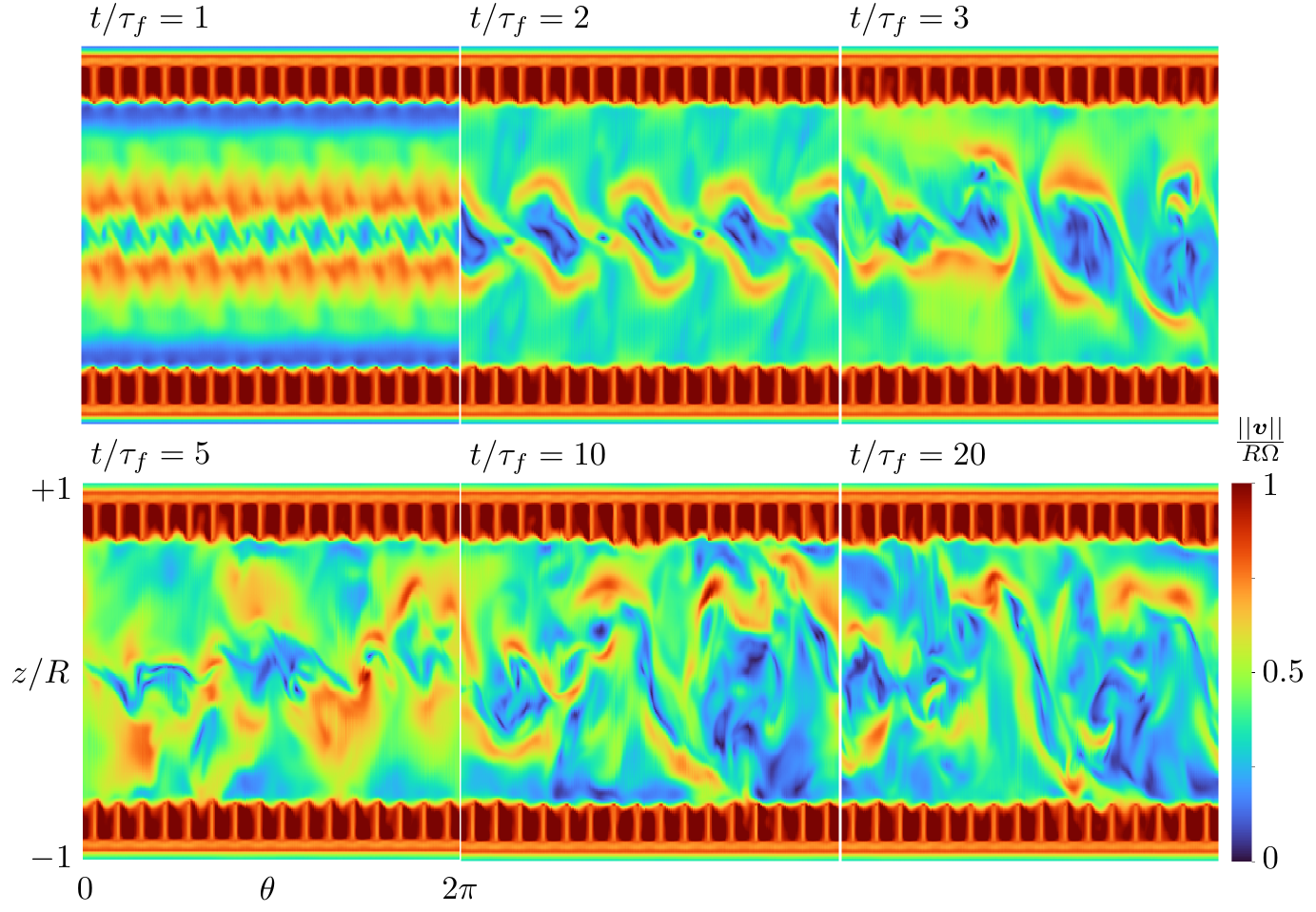}
    \caption{Isocontours of the normalized velocity magnitude at $\Rey_\Omega=2000$. The circumferential cut is taken at the radial distance $r/R=0.8$. An early instability of the shear layer grows rapidly into intense velocity fluctuations. Large vortical structures of size comparable to the disk radius traverse the shear layer. \label{fig:snap_2000}}
  \end{figure*}
  
  \begin{figure*}
    \centering
    \includegraphics[width=\linewidth]{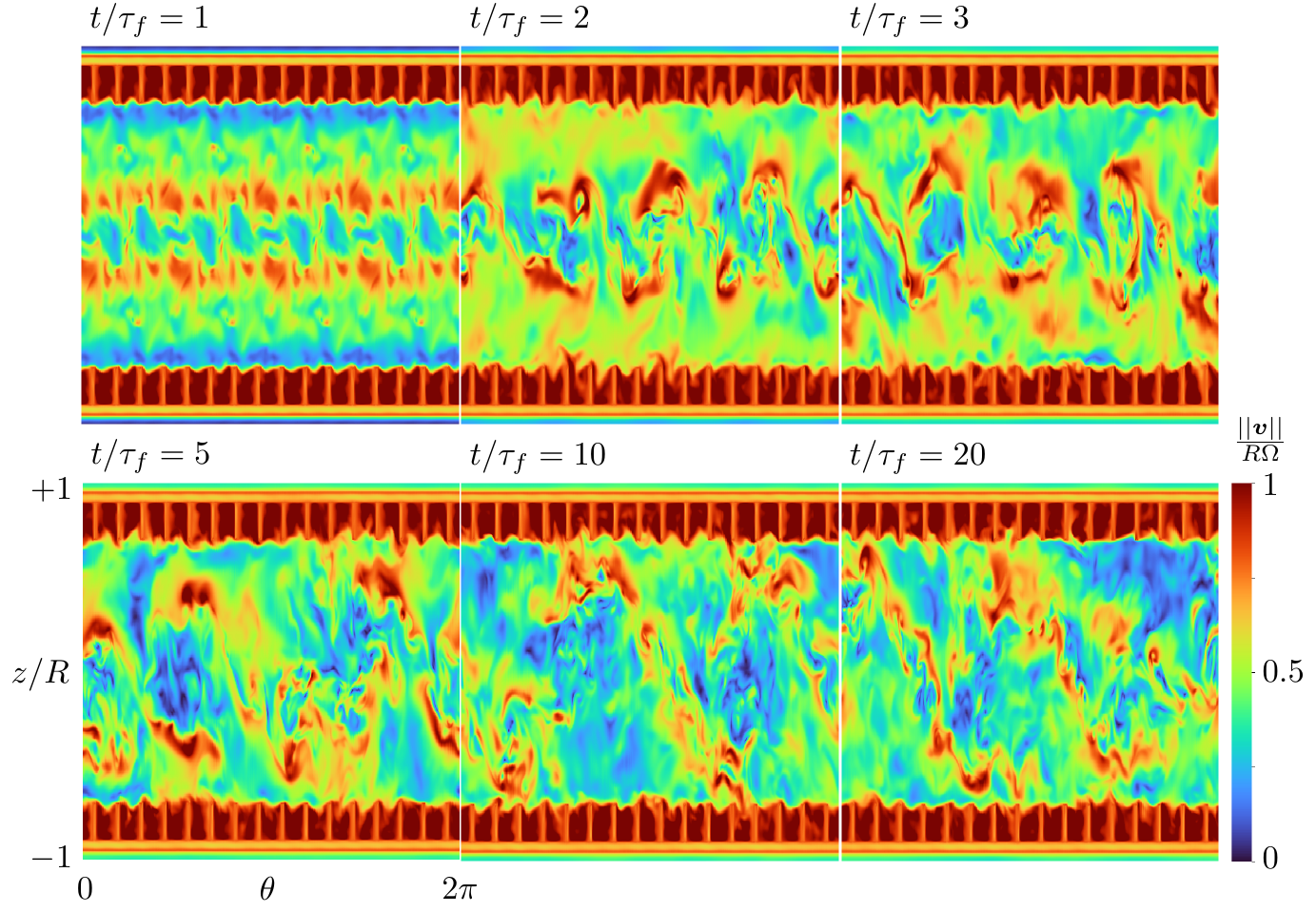}
    \caption{Isocontours of the normalized velocity magnitude at $\Rey_\Omega=4000$. The circumferential cut is taken at the radial distance $r/R=0.8$. The flow is fully turbulent and reaches a statistically stationary state in about 2 revolutions of the impellers.  \label{fig:snap_4000}}
  \end{figure*}
    
  
  We now consider the flow at the three higher Reynolds numbers, $\Rey_\Omega=2000$ and 4000. The normalized velocity magnitude at radial location $r=0.8R$ is shown in Fig. \ref{fig:snap_2000} and \ref{fig:snap_4000}. At these higher Reynolds numbers, the symmetries characterizing the low Reynolds number regimes are absent. In these two cases, the transition to turbulence occurs when an azimuthal mode $m=4$ breaks into turbulent fluctuations. The transition takes approximately 5 revolutions of the impellers. Once a statistically stationary state establishes, we observe intense velocity fluctuations sustained in the device with large vortical structures of size comparable to the disk radius traversing the shear layer. The corrugation of isocontours of the velocity magnitude display an increasing distribution of scales with increasing Reynolds number indicating a widening of the inertial range.

    \begin{figure*}
        \centering
        \includegraphics{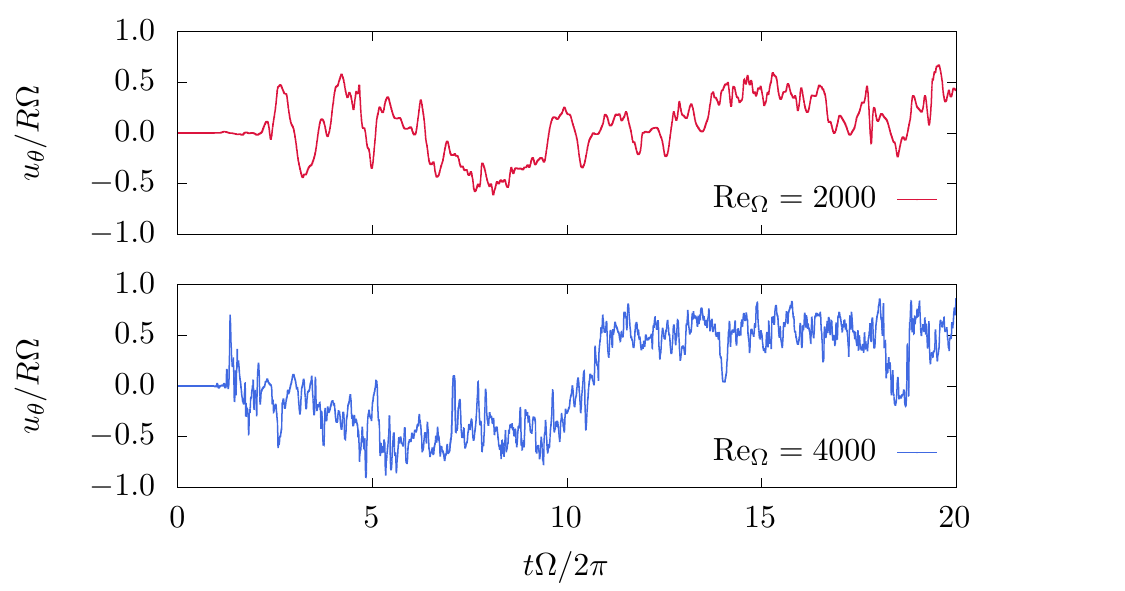}
      \caption{\color{revision1} Time series of the normalized azimuthal velocity measured at a reference point located at a radial distance $r=0.9R$ on the mid-plane. The fluctuation rms are 0.30 and 0.51 for the present DNS at $\Rey_\Omega=2000$ and $\Rey_\Omega=4000$, respectively, compared to 0.44 and 0.52 in the experiments of \citet{raveletSupercriticalTransitionTurbulence2008}.\label{fig:time_series}}
    \end{figure*}
    \begin{figure*}
      \centering
      \includegraphics[width=0.9\linewidth]{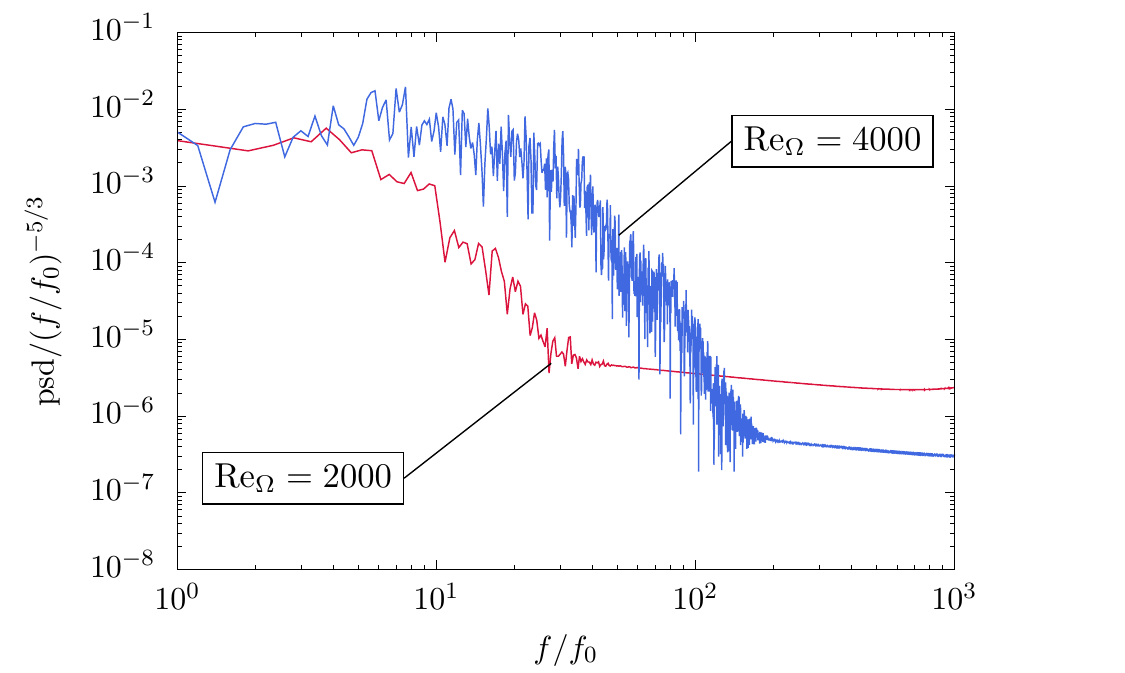}
      \caption{Compensated power spectra. Unlike the case at $\Rey=4000$, the short bandwidth of the inertial range suggests that turbulence at $\Rey_\Omega=2000$ is not fully developed. Note that $f_0=\Omega/(2\pi)$ is the frequency associated with one full revolution of the impellers.\label{fig:psd}}
    \end{figure*}

  The flow regime can be determined from analysis of the velocity fluctuations at a reference point. Following \citet{raveletSupercriticalTransitionTurbulence2008}, we measure the azimuthal velocity  at a radial location $r=0.9R$ on the mid-height plane. The time series are shown in Fig. \ref{fig:time_series}, where the velocity is normalized by the blade-tip speed $R\Omega$. It is noteworthy that the inertially-driven turbulent \VK flow achieves high turbulence intensity. \textcolor{revision1}{The root mean square (rms) of the azimuthal velocity fluctuations at the sampling location is  $u'_\theta/(R\Omega)=0.30$, 0.51 for $\Rey_\Omega=2000$ and 4000, respectively. In comparison, the experimental fit in \citep{raveletSupercriticalTransitionTurbulence2008} gives $u'_\theta/(R\Omega)=0.44$ at $\Rey_\Omega=2000$ and $u'_\theta/(R\Omega)=0.52$ at $\Rey_\Omega=4000$. The lower turbulence intensity at $\Rey_\Omega=2000$ could be due to the shorter averaging period compared to the experiments, where approximately 1000 revolutions are used.}
  The compensated power spectra of the time series are shown in Fig \ref{fig:psd}. Normalization by the $-5/3$ power-law shows the establishment of the inertial range where the curve is flat. We note that for the $\Rey_\Omega=2000$, the extent of the inertial range is smaller than a decade, which is indicative of transitional turbulence. This observation is in agreement with those of \citet{raveletSupercriticalTransitionTurbulence2008} who found that fully developed inertial turbulence is achieved for values of Reynolds number above $\sim 3300$. This regime is achieved in the present DNS at  $\Rey_\Omega=4000$ as evidenced by the inertial range extends beyond one decade in Fig. \ref{fig:psd}.

  \begin{figure*}
    \centering
        \includegraphics{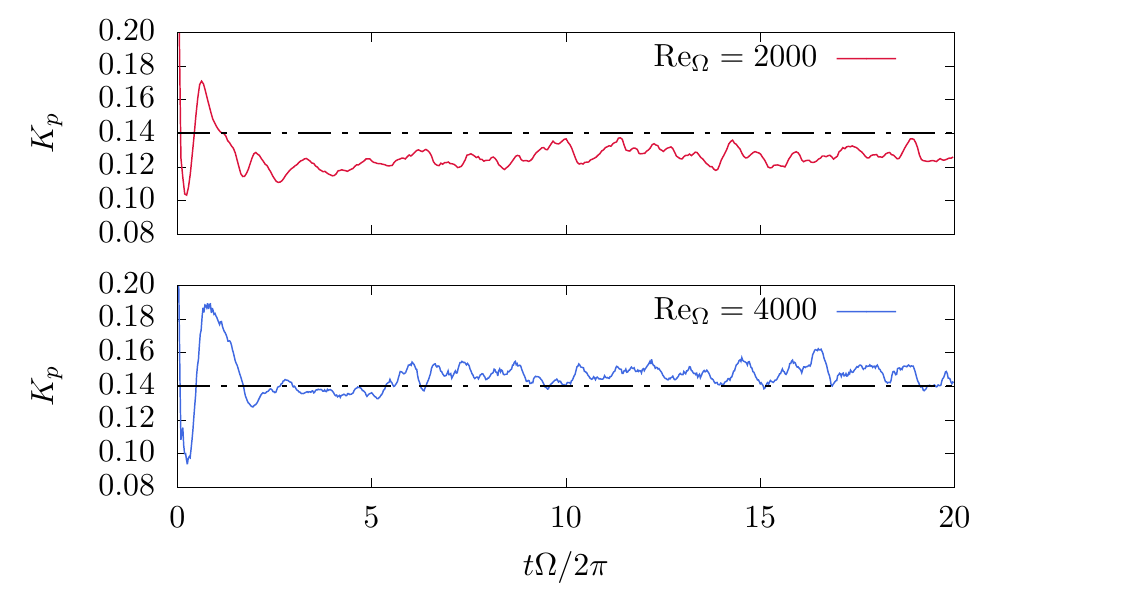}
    \caption{Non-dimensional torque. The mean stationary values are 0.1267 and 0.1472 for $\Rey_\Omega=2000$ and 4000, respectively. These values are in good agreement with the experimentally determined value $K_p=0.14$ \citep{raveletSupercriticalTransitionTurbulence2008}, represented by the dash-dotted line, in fully developed turbulence.  \label{fig:torque}}
  \end{figure*}
  Another macroscopic observable of interest is the torque exerted by the impellers in order to induce the fluid motion. In laboratory devices, torque is related to the power consumption by the motors driving the impellers. To calculate this quantity, we measure first the total power associated with the force exerted by the immersed boundaries, which is determined from the IB forcing term as
  \begin{equation}
    \mathcal{P}_\mathrm{IB}=\iiint \bm{u}\cdot\bm{F}_\mathrm{IB}dV.
  \end{equation}
  Because the impellers rotate at a constant rate $\Omega$, the relationship between power generated by the IB and torque $T$ is
  \begin{equation}
    T=\frac{\mathcal{P}_\mathrm{IB}}{\Omega}.
  \end{equation}
  According to \citet{raveletSupercriticalTransitionTurbulence2008}, the non-dimensional torque $K_p=T/(\rho R^5\Omega^2)$ reaches an asymptotic value $K_p\simeq 0.14$, independent of the Reynolds number, for $\Rey_\Omega>3300$. In order to compare with the experiments, we report the temporal evolution of $K_p$ in Fig. \ref{fig:torque}. After a transient of about 4 revolutions of the impellers, the non-dimensional torque reaches a stationary state. The mean $K_p$ establishes at 0.1267 and 0.1462 for $\Rey_\Omega=2000$ and 4000, respectively. The values found from these simulations are in excellent agreement with \citet{raveletSupercriticalTransitionTurbulence2008} since  $K_p$ is within a few percent of the experimentally determined asymptotic value.

\subsubsection{Characterization of homogeneity and isotropy in fully developed turbulence}

\begin{figure*}
  \centering
  \begin{subfigure}[b]{0.4\textwidth}
  \includegraphics[width=\linewidth,trim={40ex 15ex 36.5ex 15ex},clip]{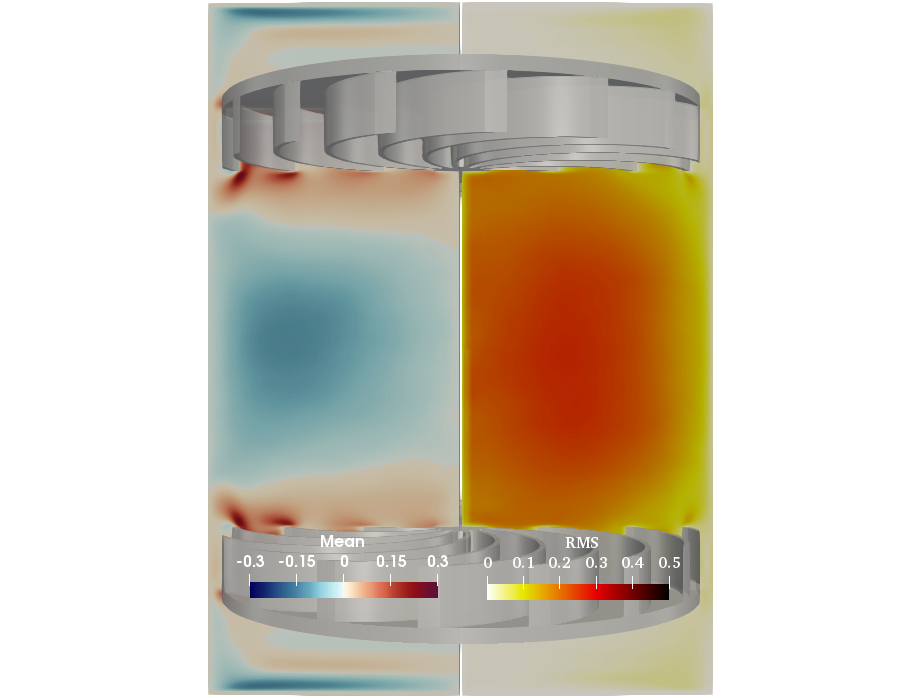}
  \caption{\label{fig:Re4000_mean_ur}}
  \end{subfigure}
  ~
  \begin{subfigure}[b]{0.55\linewidth}
  \includegraphics[width=\linewidth]{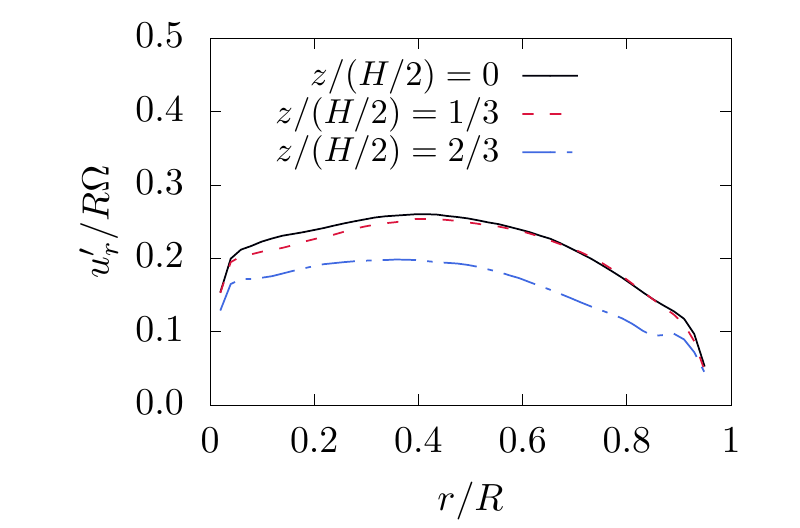}
  \caption{\label{fig:Re4000_rms_ur}}
  \end{subfigure}
  \caption{Radial velocity component. (a) Isocontours of the normalized mean (left half) and rms fluctuations (right half) of the radial velocity at $\Rey=4000$. (b) Radial profile of the normalized radial velocity fluctuations at three locations along the axis for $\Rey = 4000$.\label{fig:Re4000_ur}}
\end{figure*}

\begin{figure*}
  \centering
  \begin{subfigure}[b]{0.4\textwidth}
  \includegraphics[width=\linewidth,trim={40ex 15ex 36.5ex 15ex},clip]{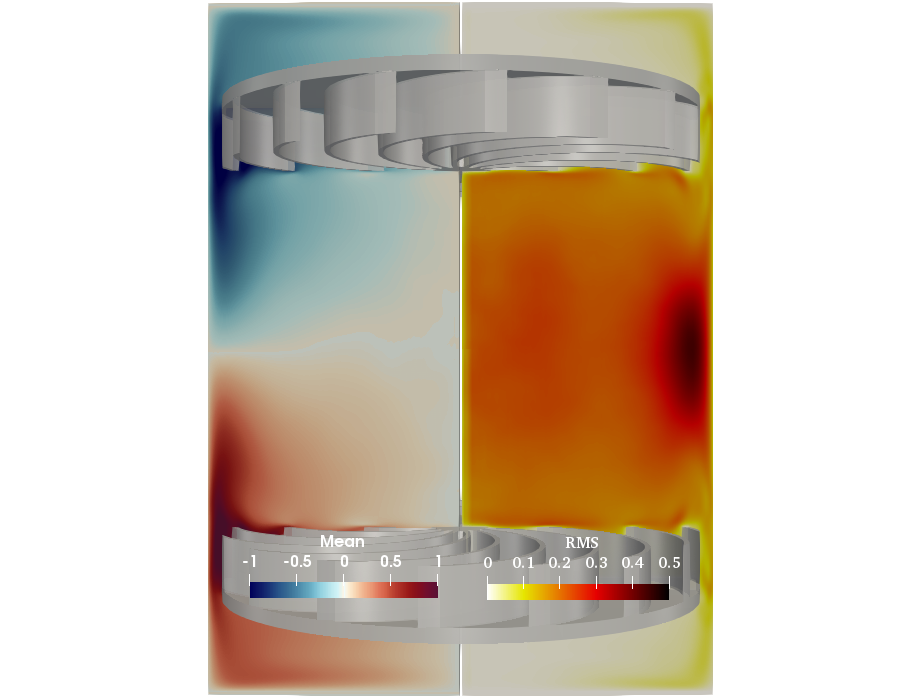}
  \caption{\label{fig:Re4000_mean_ut}}
  \end{subfigure}
  ~
  \begin{subfigure}[b]{0.55\linewidth}
  \includegraphics[width=\linewidth]{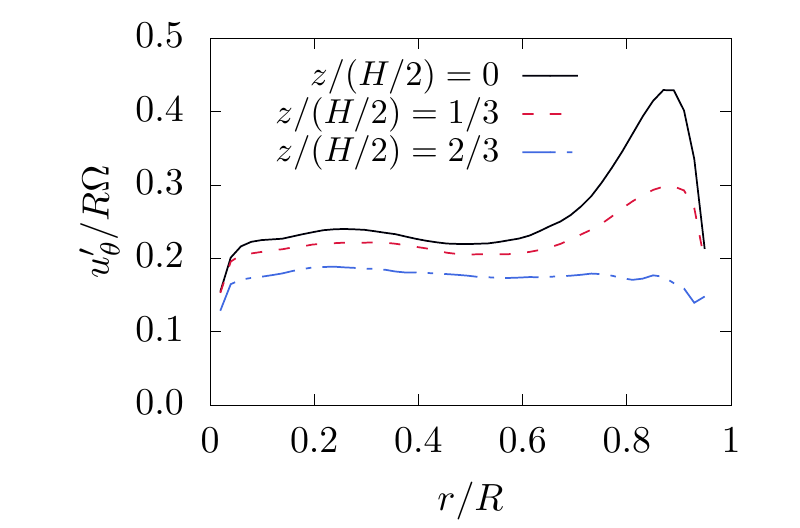}
  \caption{\label{fig:Re4000_rms_ut}}
  \end{subfigure}
  \caption{Azimuthal velocity component. (a) Isocontours of the normalized mean (left half) and rms fluctuations (right half) of the azimuthal velocity at $\Rey=4000$. (b) Radial profile of the normalized azimuthal velocity fluctuations at three locations along the axis for $\Rey = 4000$.\label{fig:Re4000_ut}}
\end{figure*}

\begin{figure*}
  \centering
  \begin{subfigure}[b]{0.4\textwidth}
  \includegraphics[width=\linewidth,trim={40ex 15ex 36.5ex 15ex},clip]{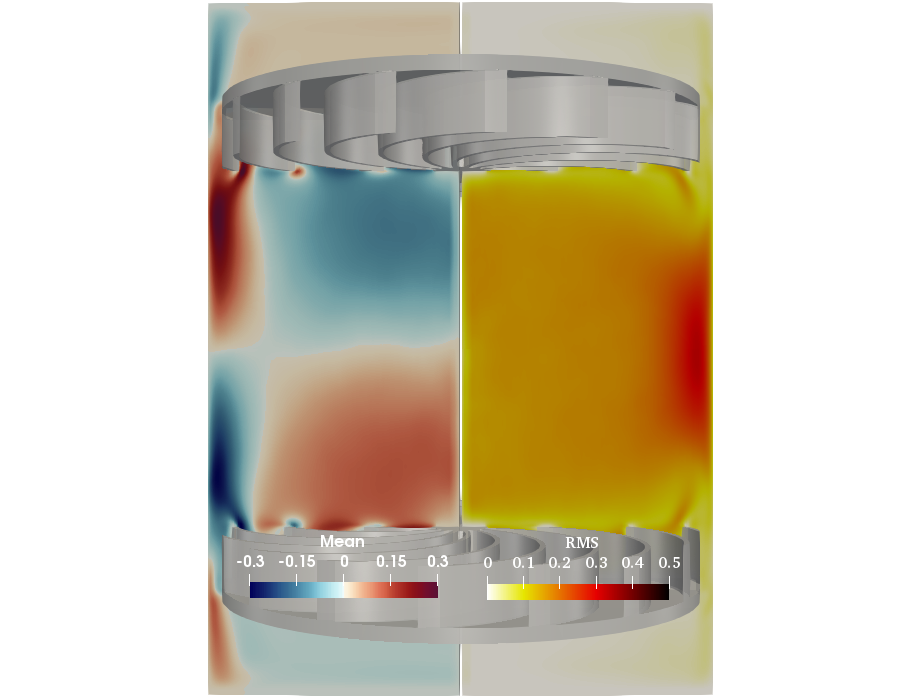}

  \caption{\label{fig:Re4000_mean_uz}}
  \end{subfigure}
  ~
  \begin{subfigure}[b]{0.55\linewidth}
  \includegraphics[width=\linewidth]{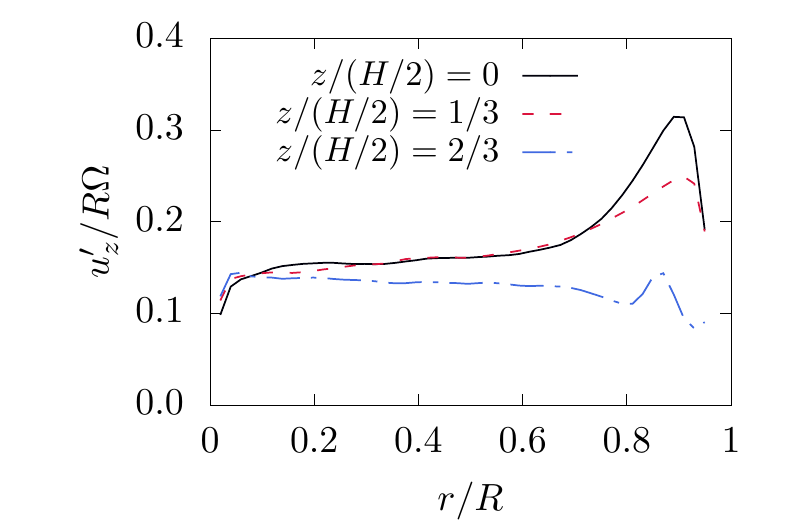}
  \caption{\label{fig:Re4000_rms_uz}}
  \end{subfigure}
  \caption{Axial velocity component. (a) Isocontours of the normalized mean (left half) and rms fluctuations (right half) of the axial velocity at $\Rey=4000$. (b) Radial profile of the normalized axial velocity fluctuations at three locations along the axis for $\Rey = 4000$.\label{fig:Re4000_uz}}
\end{figure*}

 Due to the canonical nature of the turbulent swirling \VK flow, it is worthwhile to characterize the nature of the turbulent fluctuations in the fully developed turbulence regime at $\Rey_\Omega=4000$. In particular, we seek to understand whether the turbulent fluctuations in the  central region of the flow, i.e., close to the axis and near the mid-height plane are isotropic.

In the present DNS, flow averages and fluctuating quantities are considered once the flow achieves a statistically stationary state, i.e., after 5 revolutions of the disks. Averaging is conducted from the perspective of an observer on a rotating blade  using 750 snapshots gathered over 15 rotations. Averages are obtained by grouping data at grid points at equal angles ahead of any one blade (from 0 to 45$^\circ$). Samples at different times are rotated by the corresponding angle.

Figures \ref{fig:Re4000_ur}, \ref{fig:Re4000_ut}, and \ref{fig:Re4000_uz} show the mean and the rms velocity components normalized by the blade-tip speed $R\Omega$. It is interesting to note that despite the symmetry-breaking instabilities activated at intermediate Reynolds numbers \citep{lopezInstabilityModeInteractions2002,raveletSupercriticalTransitionTurbulence2008,cortetSusceptibilityDivergencePhase2011}, the mean flow in the fully developed turbulence regime displays axial and planar symmetries. Much like in the laminar regime at $\Rey_\Omega=90$, the mean flow field consists of two toroidal cells created by fluid ejected radially outward from the blades towards the cylindrical walls, which is then redirected along the walls towards the mid-plane (Fig. \ref{fig:Re4000_mean_ur} and \ref{fig:Re4000_mean_uz}).  \citet{cortetExperimentalEvidencePhase2010} note that symmetry-breaking transitions may arise in the averaged flow as well, albeit at higher Reynolds numbers than considered here, which raises questions on the stability of the mean flow.

The presence of strong shear in the mid plane due to the two counter-rotating stacked toroidal structures generates large velocity fluctuations,
which reach about 30\% the tip speed for the azimuthal and radial components.
Shear in the boundary layers at the walls of the cylindrical enclosure generates large fluctuations in the axial and azimuthal velocity components also.
The central region of $0 \leq r/R \leq 0.5$ and $z/(H/2) < 1/3$ is of particular interest since it features small mean velocities and little spatial variation in the rms fluctuations.

\begin{figure*}
  \centering
    \includegraphics[width=0.7\linewidth]{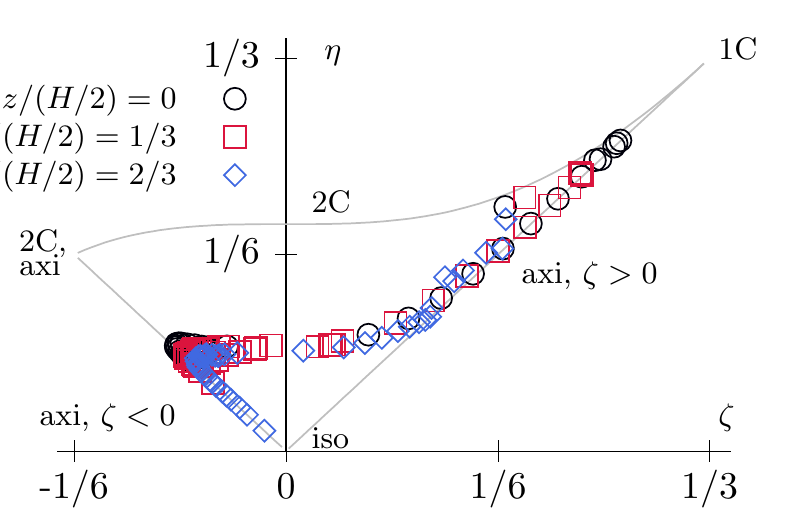}
    \caption{Lumley triangle on the plane of invariants $\zeta-\eta$ of the Reynolds stress anisotropy tensor for $\Rey = 4000$  in three planes normal to the axis.
     Turbulence in the central region is neither two-component nor truly isotropic.\label{fig:lumley_4000}}
\end{figure*}

Turbulence in the central region is not isotropic, since the axial fluctuations are smaller than the other two components, as shown in Fig. \ref{fig:Re4000_rms_ur}, \ref{fig:Re4000_rms_ut}, and \ref{fig:Re4000_rms_uz}.
The radial and azimuthal fluctuations decrease in magnitude as we move axially toward the blades.
At two thirds of the distance, the rms values of three velocity components are nearly equal and turbulence approaches isotropy.

In order to characterize the anisotropy of the velocity fluctuations, we investigate the Reynolds stress tensor $b_{ij} = \langle u_i u_j\rangle/\langle u_k u_k \rangle - \delta_{ij}/3$ and display it in the Lumley triangle \citep{lumleyReturnIsotropyHomogeneous1977} shown in Fig. \ref{fig:lumley_4000}. Here, $\mathrm{II}_b = b_{ij}b_{ji}/2$ and $\mathrm{III_b} = (b_{ij} b_{jk} b_{ki})/3$ correspond to the second and third invariants of the tensor $b_{ij}$, while $\zeta = (\mathrm{III_b}/2)^{1/3}$ and $\eta = (-\mathrm{II}_b/3)^{1/2}$ correspond to the transformed invariants.
The data is presented for various radial locations at three axial planes: $z/(H/2) = 0,1/3$ and $2/3$.
It is apparent that turbulence in the central region (black circles) is neither fully isotropic nor axisymmetric.
Moving outwards in the radial direction, the flow transitions to axisymmetric turbulence where the two eigenvalues of the anisotropy tensor are equal and smaller than the third larger eigenvalue.
The near wall region displays characteristics of a single component turbulence (labeled `1C' at the top right corner),
consistent with the presence of boundary layers near the walls of the enclosure.

\section{Conclusions}
\label{sec:conclusion}

In this study, we presented results from direct numerical simulations of the swirling \VK flow at Reynolds numbers $\Rey_\Omega=90$, 360, 2000 and 4000 in a configuration that reproduces the experiments of \citet{raveletSupercriticalTransitionTurbulence2008}. While there has been vigorous experimental work on the swirling \VK flow, DNS of this flow remain scarce. The numerical simulations presented here display qualitative and quantitative agreement across a range of flow regimes from laminar to fully developed turbulence. \textcolor{revision1}{ This shows that a straightforward implementation of the present IBM on a uniform grid is a powerful tool for the study of such impeller driven flows.}

At Reynolds numbers $\Rey_\Omega=90$, the flow consists of two toroidal cells stacked on each other. The flow is axisymmetric and planar symmetric about the mid-plane. The latter symmetry is lost at $\Rey_\Omega=360$ due to the sudden onset of a Kelvin-Helmholtz instability. An azimuthal mode $m=2$ develops on the shear layer at the mid-plane causing the distortion of the tori. These flow patterns conform closely to the dynamics identified in \citep{raveletSupercriticalTransitionTurbulence2008}  for the laminar regime, whereby successive symmetry-breaking instabilities appear with increasing Reynolds number. Analysis of time series of velocity fluctuations shows that the case at $\Rey_\Omega=2000$ is transitional, while simulations at $\Rey_\Omega=4000$ achieve fully developed turbulence. The non-dimensional torque computed from DNS matches experimental correlations remarkably well.

Results from the DNS in the fully developed regime show that the mean flow exhibits the same-symmetries as the laminar case $\Rey_\Omega=90$. This suggests that modes created by the low-Reynolds number instabilities are overshadowed by fully developed turbulence. Owing to the strong shear between the two tori, turbulent fluctuations are intense, particularly in the radial and azimuthal directions scaling as 20 to 40\% of the blade-tip velocity $R\Omega$. Using the Lumley triangle, we find that the fluctuations in the central region remain anisotropic  at $\Rey_\Omega=4000$.

The simulations are enabled by a novel immersed boundary method, which extends the approach of \citet{uhlmannImmersedBoundaryMethod2005}, and is embedded within an incompressible semi-implicit framework with a predictor-corrector step for mass conservation \citep{desjardinsHighOrderConservative2008}. The approach consists in decoupling the momentum and Eulerian IB forcing equations via operator-splitting. The latter is solved using a backward Euler scheme. Surface integrals are discretized using a triangular mesh of the surface of the immersed body. The forcing terms are computed at the centroids of the triangular faces, which are tracked in a Lagrangian reference frame for moving solids. Our strategy results in an update similar to that of \citet{uhlmannImmersedBoundaryMethod2005}, although derived differently. The robustness and stability of the methodology made the present simulations of the swirling \VK flow possible \textcolor{revision1}{with simple uniform grids}. 

\textcolor{revision1}{The use of locally refined grids as in \citep{kangDNSBuoyancydominatedTurbulent2009a} could improve the solutions near the immersed boundaries. However, for moving boundaries, such as impellers, it is not clear yet how this refinement can be achieved without incurring the same penalties found in methods using body-conformal meshes. Coupling the present IBM with overset grids could provide a way forward, and shall be investigated in future studies.}

\section*{Acknowledgment}
   Support for this research was provided in part by NSF grant CBET-1805921. The simulations were executed under the XSEDE computing grant CTS180002. The authors would like to thank Dr. B\'ereng\`ere Dubrulle for providing technical documentations on the swirling \VK flow devices. \textcolor{revision1}{We also would like to acknowledge the helpful comments of the anonymous referees}.
{\color{revision1}
\appendix
\section{Effect of grid resolution on the measured torque}
\label{sec:appendix}
To demonstrate the grid convergence of our computational method in the swirling \VK flow cases, we present the results from two auxiliary simulations at $\Rey_\Omega=4000$. Compared to the reference simulation in Tab. \ref{tab:VK_parameters}, these two additional simulations are performed on a coarser and a finer grid. The former is a uniform Cartesian grid with $256\times 342\times 256$ points, yielding a constant resolution $R/\Delta x=128$. The fine grid has  $640\times 896\times 640$ points corresponding to a resolution of $R/\Delta x=320$. Note that the simulation in Tab. \ref{tab:VK_parameters} has a size $512\times 688\times 512$ and resolution $R/\Delta x=256$.

\begin{figure} \centering
  \includegraphics[width=0.7\linewidth]{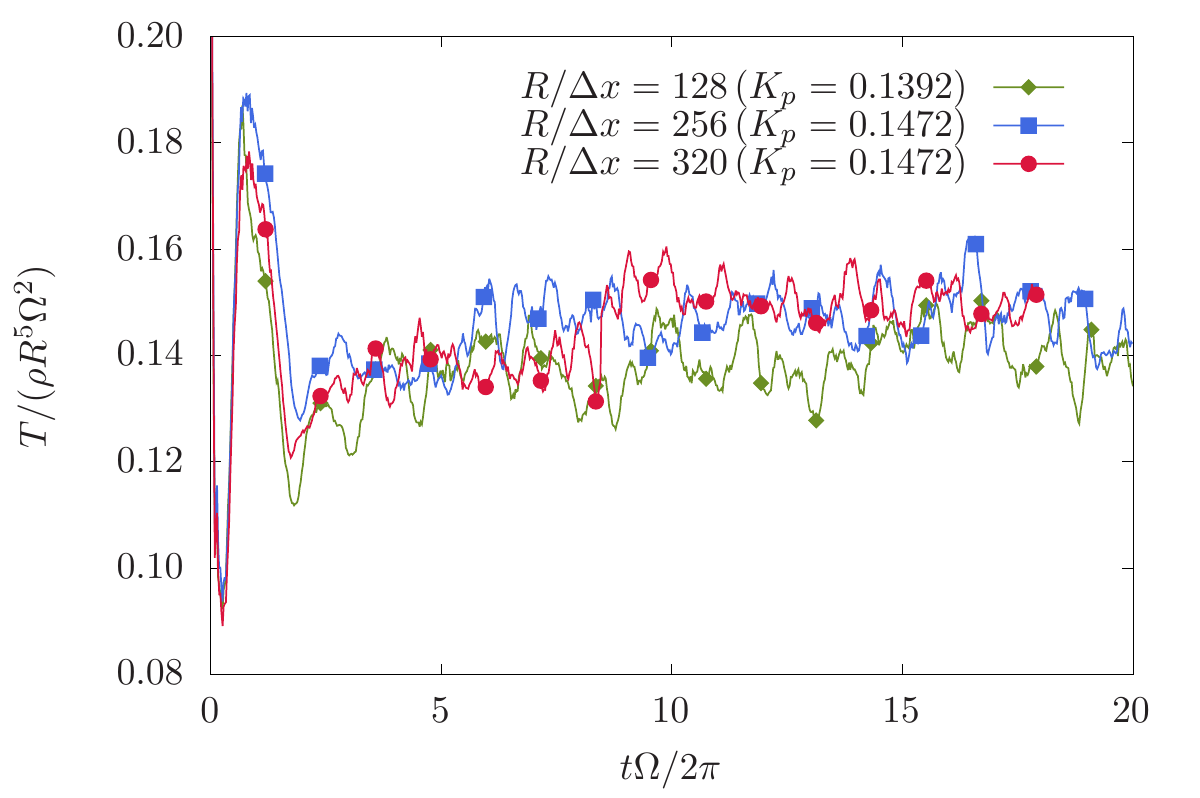}
  \caption{Evolution of the non-dimensional torque at the three mesh resolutions $R/\Delta x=$128, 256, and 320. The average non-dimensional torque, $K_p$, converges to 0.1472.\label{fig:appendix_torque}}	
\end{figure}

Figure \ref{fig:appendix_torque} shows the evolution of the non-dimensional torque from three three runs. The average non-dimensional torque, $K_p$, computed from the fifth revolution and onward, converges to 0.1472 for $R/\Delta x=256$ and 320. This convergence study shows that the fluid stresses on the impellers are well captured by resolutions $R/\delta x=256$ and beyond for swirling \VK flow at $\Rey_\Omega=4000$.
}

\bibliographystyle{plainnat}
\bibliography{references}

\begin{thebibliography}{45}
\providecommand{\natexlab}[1]{#1}
\providecommand{\url}[1]{\texttt{#1}}
\expandafter\ifx\csname urlstyle\endcsname\relax
  \providecommand{\doi}[1]{doi: #1}\else
  \providecommand{\doi}{doi: \begingroup \urlstyle{rm}\Url}\fi

\bibitem[Akselvoll(1995)]{akselvollLargeEddySimulation1995}
Knut Akselvoll.
\newblock \emph{Large Eddy Simulation of Turbulent Confined Coannular Jets and
  Turbulent Flow over a Backward Facing Step /}.
\newblock PhD thesis, Stanford University, {Stanford, CA}, 1995.

\bibitem[Balaras(2004)]{balarasModelingComplexBoundaries2004}
Elias Balaras.
\newblock Modeling complex boundaries using an external force field on fixed
  {{Cartesian}} grids in large-eddy simulations.
\newblock \emph{Computers \& Fluids}, 33\penalty0 (3):\penalty0 375--404, March
  2004.
\newblock ISSN 0045-7930.
\newblock \doi{10.1016/S0045-7930(03)00058-6}.

\bibitem[Batchelor(1951)]{batchelorNOTECLASSSOLUTIONS1951}
G.~K. Batchelor.
\newblock Note on a class of solutions of the {{Navier}}-{{Stokes}} equations
  representing steady rotationally-symmetric flow.
\newblock \emph{The Quarterly Journal of Mechanics and Applied Mathematics},
  4\penalty0 (1):\penalty0 29--41, January 1951.
\newblock ISSN 0033-5614.
\newblock \doi{10.1093/qjmam/4.1.29}.

\bibitem[Bertrand et~al.(1980)Bertrand, Couderc, and
  Angelino]{bertrandPowerConsumptionPumping1980}
J.~Bertrand, J.~P. Couderc, and H.~Angelino.
\newblock Power consumption, pumping capacity and turbulence intensity in
  baffled stirred tanks: {{Comparison}} between several turbines.
\newblock \emph{Chemical Engineering Science}, 35\penalty0 (10):\penalty0
  2157--2163, January 1980.
\newblock ISSN 0009-2509.
\newblock \doi{10.1016/0009-2509(80)85040-8}.

\bibitem[Burnishev and
  Steinberg(2014)]{burnishevTorquePressureFluctuations2014}
Yuri Burnishev and Victor Steinberg.
\newblock Torque and pressure fluctuations in turbulent von {{Karman}} swirling
  flow between two counter-rotating disks. {{I}}.
\newblock \emph{Physics of Fluids}, 26\penalty0 (5):\penalty0 055102, May 2014.
\newblock ISSN 1070-6631.
\newblock \doi{10.1063/1.4873201}.

\bibitem[Choi and Moin(1994)]{choiEffectsComputationalTime1994}
Haecheon Choi and Parviz Moin.
\newblock Effects of the {{Computational Time Step}} on {{Numerical Solutions}}
  of {{Turbulent Flow}}.
\newblock \emph{Journal of Computational Physics}, 113\penalty0 (1):\penalty0
  1--4, July 1994.
\newblock ISSN 0021-9991.
\newblock \doi{10.1006/jcph.1994.1112}.

\bibitem[Cortet et~al.(2010)Cortet, Chiffaudel, Daviaud, and
  Dubrulle]{cortetExperimentalEvidencePhase2010}
P.-P. Cortet, A.~Chiffaudel, F.~Daviaud, and B.~Dubrulle.
\newblock Experimental {{Evidence}} of a {{Phase Transition}} in a {{Closed
  Turbulent Flow}}.
\newblock \emph{Physical Review Letters}, 105\penalty0 (21), November 2010.
\newblock ISSN 0031-9007, 1079-7114.
\newblock \doi{10.1103/PhysRevLett.105.214501}.

\bibitem[Cortet et~al.(2011)Cortet, Herbert, Chiffaudel, Daviaud, Dubrulle, and
  Padilla]{cortetSusceptibilityDivergencePhase2011}
P-P Cortet, E~Herbert, A~Chiffaudel, F~Daviaud, B~Dubrulle, and V~Padilla.
\newblock Susceptibility divergence, phase transition and multistability of a
  highly turbulent closed flow.
\newblock \emph{Journal of Statistical Mechanics: Theory and Experiment},
  2011\penalty0 (07):\penalty0 P07012, July 2011.
\newblock ISSN 1742-5468.
\newblock \doi{10.1088/1742-5468/2011/07/P07012}.

\bibitem[Debue et~al.(2018)Debue, Shukla, Kuzzay, Faranda, Saw, Daviaud, and
  Dubrulle]{debueDissipationIntermittencySingularities2018}
P.~Debue, V.~Shukla, D.~Kuzzay, D.~Faranda, E.-W. Saw, F.~Daviaud, and
  B.~Dubrulle.
\newblock Dissipation, intermittency, and singularities in incompressible
  turbulent flows.
\newblock \emph{Physical Review E}, 97\penalty0 (5), May 2018.
\newblock ISSN 2470-0045, 2470-0053.
\newblock \doi{10.1103/PhysRevE.97.053101}.

\bibitem[Desjardins et~al.(2008)Desjardins, Blanquart, Balarac, and
  Pitsch]{desjardinsHighOrderConservative2008}
O.~Desjardins, G.~Blanquart, G.~Balarac, and H.~Pitsch.
\newblock High order conservative finite difference scheme for variable density
  low {{Mach}} number turbulent flows.
\newblock \emph{Journal of Computational Physics}, 227\penalty0 (15):\penalty0
  7125--7159, July 2008.
\newblock ISSN 0021-9991.
\newblock \doi{10.1016/j.jcp.2008.03.027}.

\bibitem[Dubrulle(2019)]{dubrulleKolmogorovCascades2019}
B{\'e}reng{\`e}re Dubrulle.
\newblock Beyond {{Kolmogorov}} cascades.
\newblock \emph{Journal of Fluid Mechanics}, 867, May 2019.
\newblock ISSN 0022-1120, 1469-7645.
\newblock \doi{10.1017/jfm.2019.98}.

\bibitem[Fadlun et~al.(2000)Fadlun, Verzicco, Orlandi, and
  {Mohd-Yusof}]{fadlunCombinedImmersedBoundaryFiniteDifference2000}
E.~A. Fadlun, R.~Verzicco, P.~Orlandi, and J.~{Mohd-Yusof}.
\newblock Combined {{Immersed}}-{{Boundary Finite}}-{{Difference Methods}} for
  {{Three}}-{{Dimensional Complex Flow Simulations}}.
\newblock \emph{Journal of Computational Physics}, 161\penalty0 (1):\penalty0
  35--60, June 2000.
\newblock ISSN 0021-9991.
\newblock \doi{10.1006/jcph.2000.6484}.

\bibitem[Kang et~al.(2009)Kang, Iaccarino, and
  Ham]{kangDNSBuoyancydominatedTurbulent2009a}
Seongwon Kang, Gianluca Iaccarino, and Frank Ham.
\newblock {{DNS}} of buoyancy-dominated turbulent flows on a bluff body using
  the immersed boundary method.
\newblock \emph{Journal of Computational Physics}, 228\penalty0 (9):\penalty0
  3189--3208, May 2009.
\newblock ISSN 0021-9991.
\newblock \doi{10.1016/j.jcp.2008.12.037}.

\bibitem[K{\'a}rm{\'a}n(1921)]{karmanUberLaminareUnd1921}
Th~V. K{\'a}rm{\'a}n.
\newblock \"uber laminare und turbulente {{Reibung}}.
\newblock \emph{ZAMM - Journal of Applied Mathematics and Mechanics /
  Zeitschrift f\"ur Angewandte Mathematik und Mechanik}, 1\penalty0
  (4):\penalty0 233--252, 1921.
\newblock ISSN 1521-4001.
\newblock \doi{10.1002/zamm.19210010401}.

\bibitem[Kim and Choi(2006)]{kimImmersedBoundaryMethod2006}
Dokyun Kim and Haecheon Choi.
\newblock Immersed boundary method for flow around an arbitrarily moving body.
\newblock \emph{Journal of Computational Physics}, 212\penalty0 (2):\penalty0
  662--680, March 2006.
\newblock ISSN 0021-9991.
\newblock \doi{10.1016/j.jcp.2005.07.010}.

\bibitem[Kim et~al.(2001)Kim, Kim, and
  Choi]{kimImmersedBoundaryFiniteVolumeMethod2001}
Jungwoo Kim, Dongoo Kim, and Haecheon Choi.
\newblock An {{Immersed}}-{{Boundary Finite}}-{{Volume Method}} for
  {{Simulations}} of {{Flow}} in {{Complex Geometries}}.
\newblock \emph{Journal of Computational Physics}, 171\penalty0 (1):\penalty0
  132--150, July 2001.
\newblock ISSN 0021-9991.
\newblock \doi{10.1006/jcph.2001.6778}.

\bibitem[Kreuzahler et~al.(2014)Kreuzahler, Schulz, Homann, Ponty, and
  Grauer]{kreuzahlerNumericalStudyImpellerdriven2014}
S.~Kreuzahler, D.~Schulz, H.~Homann, Y.~Ponty, and R.~Grauer.
\newblock Numerical study of impeller-driven von {{K\'arm\'an}} flows via a
  volume penalization method.
\newblock \emph{New Journal of Physics}, 16\penalty0 (10):\penalty0 103001,
  October 2014.
\newblock ISSN 1367-2630.
\newblock \doi{10.1088/1367-2630/16/10/103001}.

\bibitem[Kuzzay et~al.(2015)Kuzzay, Faranda, and
  Dubrulle]{kuzzayGlobalVsLocal2015}
Denis Kuzzay, Davide Faranda, and B{\'e}reng{\`e}re Dubrulle.
\newblock Global vs local energy dissipation: {{The}} energy cycle of the
  turbulent von {{K\'arm\'an}} flow.
\newblock \emph{Physics of Fluids}, 27\penalty0 (7):\penalty0 075105, July
  2015.
\newblock ISSN 1070-6631.
\newblock \doi{10.1063/1.4923750}.

\bibitem[Lai and Peskin(2000)]{laiImmersedBoundaryMethod2000}
Ming-Chih Lai and Charles~S. Peskin.
\newblock An {{Immersed Boundary Method}} with {{Formal Second}}-{{Order
  Accuracy}} and {{Reduced Numerical Viscosity}}.
\newblock \emph{Journal of Computational Physics}, 160\penalty0 (2):\penalty0
  705--719, May 2000.
\newblock ISSN 00219991.
\newblock \doi{10.1006/jcph.2000.6483}.

\bibitem[Liu et~al.(1998)Liu, Zheng, and
  Sung]{liuPreconditionedMultigridMethods1998}
C.~Liu, X.~Zheng, and C.~H. Sung.
\newblock Preconditioned {{Multigrid Methods}} for {{Unsteady Incompressible
  Flows}}.
\newblock \emph{Journal of Computational Physics}, 139\penalty0 (1):\penalty0
  35--57, January 1998.
\newblock ISSN 0021-9991.
\newblock \doi{10.1006/jcph.1997.5859}.

\bibitem[Lopez et~al.(2002)Lopez, Hart, Marques, Kittelman, and
  Shen]{lopezInstabilityModeInteractions2002}
J.~M. Lopez, J.~E. Hart, F.~Marques, S.~Kittelman, and J.~Shen.
\newblock Instability and mode interactions in a differentially driven rotating
  cylinder.
\newblock \emph{Journal of Fluid Mechanics}, 462:\penalty0 383--409, July 2002.
\newblock ISSN 1469-7645, 0022-1120.
\newblock \doi{10.1017/S0022112002008649}.

\bibitem[Lu and Dalton(1996)]{luCalculationTimingVortex1996}
X.~Y. Lu and C.~Dalton.
\newblock Calculation of the timing of vortex formation from an oscillating
  cylinder.
\newblock \emph{Journal of Fluids and Structures}, 10\penalty0 (5):\penalty0
  527--541, July 1996.
\newblock ISSN 0889-9746.
\newblock \doi{10.1006/jfls.1996.0035}.

\bibitem[Lumley and Newman(1977)]{lumleyReturnIsotropyHomogeneous1977}
John~L. Lumley and Gary~R. Newman.
\newblock The return to isotropy of homogeneous turbulence.
\newblock \emph{Journal of Fluid Mechanics}, 82\penalty0 (1):\penalty0
  161--178, August 1977.
\newblock ISSN 1469-7645, 0022-1120.
\newblock \doi{10.1017/S0022112077000585}.

\bibitem[Maurer et~al.(1994)Maurer, Tabeling, and
  Zocchi]{maurerStatisticsTurbulenceTwo1994}
J.~Maurer, P.~Tabeling, and G.~Zocchi.
\newblock Statistics of {{Turbulence}} between {{Two Counterrotating Disks}} in
  {{Low}}-{{Temperature Helium Gas}}.
\newblock \emph{EPL (Europhysics Letters)}, 26\penalty0 (1):\penalty0 31, 1994.
\newblock ISSN 0295-5075.
\newblock \doi{10.1209/0295-5075/26/1/006}.

\bibitem[Mittal and Iaccarino(2005)]{mittalImmersedBoundaryMethods2005}
Rajat Mittal and Gianluca Iaccarino.
\newblock Immersed {{Boundary Methods}}.
\newblock \emph{Annual Review of Fluid Mechanics}, 37\penalty0 (1):\penalty0
  239--261, 2005.
\newblock \doi{10.1146/annurev.fluid.37.061903.175743}.

\bibitem[Monchaux et~al.(2008)Monchaux, Cortet, Chavanis, Chiffaudel, Daviaud,
  Diribarne, and
  Dubrulle]{monchauxFluctuationDissipationRelationsStatistical2008}
Romain Monchaux, Pierre-Philippe Cortet, Pierre-Henri Chavanis, Arnaud
  Chiffaudel, Fran{\c c}ois Daviaud, Pantxo Diribarne, and B{\'e}reng{\`e}re
  Dubrulle.
\newblock Fluctuation-{{Dissipation Relations}} and {{Statistical
  Temperatures}} in a {{Turbulent}} von {{K\'arm\'an Flow}}.
\newblock \emph{Physical Review Letters}, 101\penalty0 (17), October 2008.
\newblock ISSN 0031-9007, 1079-7114.
\newblock \doi{10.1103/PhysRevLett.101.174502}.

\bibitem[Nicolaou et~al.(2015)Nicolaou, Jung, and
  Zaki]{nicolaouRobustDirectforcingImmersed2015}
L.~Nicolaou, S.~Y. Jung, and T.~A. Zaki.
\newblock A robust direct-forcing immersed boundary method with enhanced
  stability for moving body problems in curvilinear coordinates.
\newblock \emph{Computers \& Fluids}, 119:\penalty0 101--114, September 2015.
\newblock ISSN 0045-7930.
\newblock \doi{10.1016/j.compfluid.2015.06.030}.

\bibitem[Nore et~al.(2003)Nore, Tuckerman, Daube, and
  Xin]{noreRatioModeInteraction2003}
C.~Nore, L.~S. Tuckerman, O.~Daube, and S.~Xin.
\newblock The 1[ratio]2 mode interaction in exactly counter-rotating von
  {{K\'arm\'an}} swirling flow.
\newblock \emph{Journal of Fluid Mechanics}, 477, February 2003.
\newblock ISSN 0022-1120, 1469-7645.
\newblock \doi{10.1017/S0022112002003075}.

\bibitem[Nore et~al.(2004)Nore, Tartar, Daube, and
  Tuckerman]{noreSurveyInstabilityThresholds2004}
C.~Nore, M.~Tartar, O.~Daube, and L.~S. Tuckerman.
\newblock Survey of instability thresholds of flow between exactly
  counter-rotating disks.
\newblock \emph{Journal of Fluid Mechanics}, 511:\penalty0 45--65, July 2004.
\newblock ISSN 1469-7645, 0022-1120.
\newblock \doi{10.1017/S0022112004008559}.

\bibitem[Nore et~al.(2018)Nore, Quiroz, Cappanera, and
  Guermond]{noreNumericalSimulationKarman2018}
C.~Nore, D.~Castanon Quiroz, L.~Cappanera, and J.-L. Guermond.
\newblock Numerical simulation of the von {{K\'arm\'an}} sodium dynamo
  experiment.
\newblock \emph{Journal of Fluid Mechanics}, 854:\penalty0 164--195, November
  2018.
\newblock ISSN 0022-1120, 1469-7645.
\newblock \doi{10.1017/jfm.2018.582}.

\bibitem[Odier et~al.(1998)Odier, Pinton, and
  Fauve]{odierAdvectionMagneticField1998}
P.~Odier, J.-F. Pinton, and S.~Fauve.
\newblock Advection of a magnetic field by a turbulent swirling flow.
\newblock \emph{Physical Review E}, 58\penalty0 (6):\penalty0 7397--7401,
  December 1998.
\newblock ISSN 1063-651X, 1095-3787.
\newblock \doi{10.1103/PhysRevE.58.7397}.

\bibitem[Peskin(1972)]{peskinFlowPatternsHeart1972}
Charles~S Peskin.
\newblock Flow patterns around heart valves: {{A}} numerical method.
\newblock \emph{Journal of Computational Physics}, 10\penalty0 (2):\penalty0
  252--271, October 1972.
\newblock ISSN 0021-9991.
\newblock \doi{10.1016/0021-9991(72)90065-4}.

\bibitem[Peskin(2002)]{peskinImmersedBoundaryMethod2002}
Charles~S. Peskin.
\newblock The immersed boundary method.
\newblock \emph{Acta Numerica}, 11:\penalty0 479--517, January 2002.
\newblock ISSN 1474-0508, 0962-4929.
\newblock \doi{10.1017/S0962492902000077}.

\bibitem[Pierce(2001)]{pierceProgressVariableApproachLargeEddy2001}
C.~D. Pierce.
\newblock \emph{Progress-{{Variable Approach}} for {{Large}}-{{Eddy
  Simulation}} of {{Turbulent Combustion}}}.
\newblock PhD thesis, Stanford University, {Stanford, CA}, 2001.

\bibitem[Pierce and Moin(2004)]{pierceProgressvariableApproachLargeeddy2004}
Charles~D. Pierce and Parviz Moin.
\newblock Progress-variable approach for large-eddy simulation of non-premixed
  turbulent combustion.
\newblock \emph{Journal of Fluid Mechanics}, 504:\penalty0 73--97, April 2004.
\newblock ISSN 1469-7645, 0022-1120.
\newblock \doi{10.1017/S0022112004008213}.

\bibitem[Ravelet et~al.(2005)Ravelet, Chiffaudel, Daviaud, and
  L{\'e}orat]{raveletExperimentalKarmanDynamo2005}
F.~Ravelet, A.~Chiffaudel, F.~Daviaud, and J.~L{\'e}orat.
\newblock Toward an experimental von {{K\'arm\'an}} dynamo: {{Numerical}}
  studies for an optimized design.
\newblock \emph{Physics of Fluids}, 17\penalty0 (11):\penalty0 117104, November
  2005.
\newblock ISSN 1070-6631.
\newblock \doi{10.1063/1.2130745}.

\bibitem[Ravelet(2005)]{raveletBifurcationsGlobalesHydrodynamiques2005}
Florent Ravelet.
\newblock \emph{{Bifurcations globales hydrodynamiques et
  magnetohydrodynamiques dans un ecoulement de von Karman turbulent}}.
\newblock PhD thesis, Ecole Polytechnique X, September 2005.

\bibitem[Ravelet et~al.(2008)Ravelet, Chiffaudel, and
  Daviaud]{raveletSupercriticalTransitionTurbulence2008}
Florent Ravelet, Arnaud Chiffaudel, and Fran{\c c}ois Daviaud.
\newblock Supercritical transition to turbulence in an inertially driven von
  {{K\'arm\'an}} closed flow.
\newblock \emph{Journal of Fluid Mechanics}, 601:\penalty0 339--364, April
  2008.
\newblock ISSN 1469-7645, 0022-1120.
\newblock \doi{10.1017/S0022112008000712}.

\bibitem[Roma et~al.(1999)Roma, Peskin, and
  Berger]{romaAdaptiveVersionImmersed1999}
Alexandre~M Roma, Charles~S Peskin, and Marsha~J Berger.
\newblock An {{Adaptive Version}} of the {{Immersed Boundary Method}}.
\newblock \emph{Journal of Computational Physics}, 153\penalty0 (2):\penalty0
  509--534, August 1999.
\newblock ISSN 0021-9991.
\newblock \doi{10.1006/jcph.1999.6293}.

\bibitem[Sch{\"a}fer et~al.(1996)Sch{\"a}fer, Turek, Durst, Krause, and
  Rannacher]{schaferBenchmarkComputationsLaminar1996}
M.~Sch{\"a}fer, S.~Turek, F.~Durst, E.~Krause, and R.~Rannacher.
\newblock Benchmark {{Computations}} of {{Laminar Flow Around}} a {{Cylinder}}.
\newblock In Ernst~Heinrich Hirschel, editor, \emph{Flow {{Simulation}} with
  {{High}}-{{Performance Computers II}}: {{DFG Priority Research Programme
  Results}} 1993\textendash 1995}, Notes on {{Numerical Fluid Mechanics}}
  ({{NNFM}}), pages 547--566. {Vieweg+Teubner Verlag}, {Wiesbaden}, 1996.
\newblock ISBN 978-3-322-89849-4.
\newblock \doi{10.1007/978-3-322-89849-4_39}.

\bibitem[Uhlmann(2005)]{uhlmannImmersedBoundaryMethod2005}
Markus Uhlmann.
\newblock An immersed boundary method with direct forcing for the simulation of
  particulate flows.
\newblock \emph{Journal of Computational Physics}, 209\penalty0 (2):\penalty0
  448--476, November 2005.
\newblock ISSN 0021-9991.
\newblock \doi{10.1016/j.jcp.2005.03.017}.

\bibitem[Vanella and
  Balaras(2009)]{vanellaMovingleastsquaresReconstructionEmbeddedboundary2009}
Marcos Vanella and Elias Balaras.
\newblock A moving-least-squares reconstruction for embedded-boundary
  formulations.
\newblock \emph{Journal of Computational Physics}, 228\penalty0 (18):\penalty0
  6617--6628, October 2009.
\newblock ISSN 0021-9991.
\newblock \doi{10.1016/j.jcp.2009.06.003}.

\bibitem[Williamson(1989)]{williamsonObliqueParallelModes1989}
C.~H.~K. Williamson.
\newblock Oblique and parallel modes of vortex shedding in the wake of a
  circular cylinder at low {{Reynolds}} numbers.
\newblock \emph{Journal of Fluid Mechanics}, 206:\penalty0 579--627, September
  1989.
\newblock ISSN 0022-1120, 1469-7645.
\newblock \doi{10.1017/S0022112089002429}.

\bibitem[Yang and Balaras(2006)]{yangEmbeddedboundaryFormulationLargeeddy2006}
Jianming Yang and Elias Balaras.
\newblock An embedded-boundary formulation for large-eddy simulation of
  turbulent flows interacting with moving boundaries.
\newblock \emph{Journal of Computational Physics}, 215\penalty0 (1):\penalty0
  12--40, June 2006.
\newblock ISSN 0021-9991.
\newblock \doi{10.1016/j.jcp.2005.10.035}.

\bibitem[Zandbergen and Dijkstra(1987)]{zandbergenKarmanSwirlingFlows1987}
P~J Zandbergen and D~Dijkstra.
\newblock Von {{Karman Swirling Flows}}.
\newblock \emph{Annual Review of Fluid Mechanics}, 19\penalty0 (1):\penalty0
  465--491, 1987.
\newblock \doi{10.1146/annurev.fl.19.010187.002341}.

\end{thebibliography}
\end{document}